A.V. Novikov-Borodin

# ONTOLOGICAL SYSTEMS IN COGNITION

**Abstract.** There are investigated the generalized methods of cognition of the Existing, i.e. everything that is able to influence to the cognizer, and everything differed from the Existing is postulated as indistinguishable from the non-existing and incognizable. The traditional methods of cognition, based on the identification of the objects of the surrounding world, take into account not all influences, because are limited by the nature, understood as the ontology of the subject of cognition. The ontology is determined on the level of basic notions and definitions, so even if objects from the representation systems different by ontology are identified, they look undetermined, uncertain and paradoxical, exactly as, for example, the quantum mechanical objects or cosmological dark matter and energy. The representation systems with different ontology are incompatible with each other on the level of basic notions and cannot be unified on frames of one consistent system or theory. The concept of incompatible representations unifies on the level exceeded the ontological frames the different by ontology representation systems and incompatible approaches, such as relativistic and quantum ones. The concept gives a possibility to eliminate the principal contradictions of the system approach in scientific cognition, so solves the problem of understanding in science. The generalized methods of cognition are by definition on the foundations of different representation systems: scientific, philosophical, religious and social ones, what gives a possibility of general analysis and revision of their foundations. The interconnections of some physical theories, the philosophical, religious and social systems are briefly analyzed from this point of view. There are briefly considered the multi-polar picture of the world, which the concept of incompatible representations leads to.

## Introduction

By trying "*to put together by means of a system all cognizable phenomena of our world*" (A. Einstein), a science penetrates deeper and deeper inside matter, looks further and further in the universe. It is difficult to overestimate its achievements, but some properties and natural laws discovered experimentally and even successfully used, in no way cannot be embedded in one consistent system of representations, i.e. the scientific picture of the world, and destroy "*the entire image of the subject of scientific researches*" ([1]: the scientific picture of the world): or space and time occur relative; or particles behave like the waves [2-5]; or matter is attracted by invisible force of unknown nature in galaxies and, vice versa, the galaxies are flying away with acceleration [6,7]. It is thus found out that invisible and strange dark matter and dark energy of unknown nature, which are responsible for such attractions and accelerations, are more than 95 % of substance in the universe, and all previous centuries the physics had investigated just a small part of 'normal' matter.

After many unsuccessful efforts to put together such observed phenomena into one consistent system, for example, by means of unifying the relativistic and quantum approaches, there was established the opinion that discovered objects and laws are so far from the world of everyday phenomena that the efforts to understand them, by putting in one system with 'ordinary' objects and notions are impossible. Such paradigm negates the `***understanding***' in physics and leads to mathematical formalism, in which the possibility of critical analysis of introduced notions disappears, there is dramatically increasing the probability of speculations in science and its transformation into farce. The situation comes to vanishing of a notion of matter itself, so a subject and an aim of scientific knowledge is being lost: *"There have not remained any of classical definitions of matter in modern physics. However both philosophy and physics prefer to bypass this uncertain and dark notion, replacing it by others – space-time, chaos, system, etc."* ([1]: matter). Trying to avoid the consideration of such questions and considering them as philosophical ones, the modern physics is moving away from its fundamental advantages – the systematic approach and the understanding, without which the further progress in science becomes problematic.

Probably, the systematic approach in science has really come to its end? Dialectics, for example, as *"a logical form and general method of the reflexive theoretical thinking, having as its subject the contradictions of its conceivable content"* ([1]: dialectics), considers the contradictions as an inherent part of cognition, but they are incompatible with consistent systematic approach, which is in foundations of scientific theories and methods of cognition. The situation looks hopeless. However, it is not a first time when science meets problems, which looked insuperable, but it had occurred in a result, what exactly our delusions were insuperable, and just another time we are at the stage of reexamination of fundamentals and at the transfer to new possibilities.

## 1. Traditional and generalized processes of cognition

In the traditional for science materialistic understanding of the process of cognition the **subject** of cognition identifies the **objects** of his surrounding world through their properties.

> ***The process of cognition*** (*traditional*) is a process of identification by the subject of cognition the objects of the world surrounding him and their properties.

The set of identified objects forms the ***representations*** of the subject about the surrounding world perceived by him or his ***system of representations*** if they are systematized by some ***rules***. The system is ***consistent*** if these rules are valid everywhere in it. To build the *consistent* representation system is traditionally considered as the aim of the scientific cognition.

Such approach to cognition leads to a number of fundamental problems, interconnected with incompleteness of identified object images, with their unconformity with 'real objects', with limitations of methods of cognition, and, as a consequence, with a general cognoscibility of the world. Without going beforehand into discussions on these questions, we will start from the following *statement*:

> ***Postulate*** (*of existence and cognoscibility*). Everything, which, somehow or other, directly or indirectly, may influence to the cognize, exists for him and may be cognized by him.

Strictly speaking, 'Everything, which exists for the cognizer' differs from 'Everything, which may be in general', because 'Something, which cannot influence at all to the cognizer' may well exist. However, it will be 'indistinguishable from non-Existing' for him and, therefore, cannot be cognized by him. Thus, '***Everything***, which may be in general' ($\mathcal{W}$) for the cognizer $\wp$ consists from 'Everything, which can somehow influence him' – its ***Existing*** $\mathcal{E}_\wp$ and 'Something, which cannot influence him', i.e. the logical negotiation of the Existing – the ***non-Existing*** $\mathcal{N}_\wp = (\neg \mathcal{E}_\wp)$:

$$\mathcal{W} = \mathcal{E}_\wp \vee (\neg \mathcal{E}_\wp) = \mathcal{E}_\wp \vee \mathcal{N}_\wp. \tag{1}$$

The cognizer is an inherent part of his process of cognition, so exists for him and is by definition a part of his Existing. Thus, the possibility of separation of the Existing onto parts is initially installed by the process of cognition itself, which separates the cognizer from the Existing as its part:

$$\wp \subset \mathcal{E}_\wp. \tag{2}$$

Being a part of his Existing, the cognizer by definition has to influence to himself, also influencing to that part of his Existing, which is himself. Thus, the cognizer ***interacts*** with his Existing through himself and is ***interconnected*** with his Existing through these interactions. The interactions of the Existing with its parts will be determined as the general ***process of cognition***.

> ***The process of cognition*** (*generalized*) is a process of interaction of the Existing with its part.

Unlike the cognizer in the generalized process of cognition the subject of cognition in traditional process does not simply interacts with the parts of the Existing, but *identifies* them as ***objects***. Usually, the identification is associated with the conscious activity, and it causes additional difficulties with the definition of the subject of cognition. There are no such problems with the cognizer, because every part of the Existing can interact with it, so every part may be in general considered as a cognizer. This way, the changes, which the part of the Existing undergoes during the interactions with other parts, may be well interpreted as a *result* of cognition. In such approach, even inanimate objects may be considered as the cognizers. For example, the evidences in criminalistics or the artefacts in archeology may be well considered as the witnesses of events, also as the visual or audio information may be considered as testimonies of electromagnetic fields and sound waves. Thus, the traditional process of cognition is a particular case of the generalized one.

> ***Theorem*** (*the unity of the Existing*). The Existing is interconnected and closed for all its parts, it is one and the only one for all cognizers in it.

■ So as the cognizer is directly or indirectly interconnected with any part of the Existing, all parts are connected with each other at least through him, so the Existing is *connected*. Everything connected with the cognizer's Existing is indirectly connected with him and by definition is a part of his Existing, so all parts of the Existing are interconnected only with each other and they do not have any other interconnections. Thus, the Existing is *closed* for all its parts, including all cognizers in it.

Let the cognizers $p, q,...$ with their Existings $\mathcal{E}_p, \mathcal{E}_q, ...$: $p \subset \mathcal{E}_p$, $q \subset \mathcal{E}_q,...$, are parts of the Existing $\mathcal{E}$: $p, q, ... \subset \mathcal{E}$. As far as $\mathcal{E}$ is connected, the cognizers $p, q,...$ are interconnected and their Existings are also interconnected with each other through them, so are parts of the Existing $\mathcal{E}$: $\mathcal{E}_p, \mathcal{E}_q, ... \subseteq \mathcal{E}$. From other side, the Existing $\mathcal{E}$ is interconnected with the cognizers $p, q,...$ and has to be a part of the Existing of each of them: $\mathcal{E} \subseteq \mathcal{E}_p, \mathcal{E}_q, ....$ From conditions $\mathcal{E} \subseteq \mathcal{E}_p, \mathcal{E}_q, ... \subseteq \mathcal{E}$ one can get:

$$\mathcal{E}_p \equiv \mathcal{E}_q \equiv \cdots \equiv \mathcal{E}, \forall p, q, ... \subset \mathcal{E}, \qquad (3)$$

so the Existing is *one and the only one* for all cognizers in it. ■

Thus, for any cognizer $p, q,...$ from the Existing $\mathcal{E}$, everything differed from their common Existing $\mathcal{E}$ is indistinguishable from the non-Existing: $\mathcal{N} = (\neg \mathcal{E})$, so taking into account (3), the expression (1) may be rewritten as:

$$\mathcal{W} = \mathcal{E} \vee (\neg \mathcal{E}) = \mathcal{E} \vee \mathcal{N}. \qquad (4)$$

Further, if other is not specified, there will be considered the only Existing $\mathcal{E}$ and cognizers only in it.

## 2. Incompatible representations

Let some parts of the Existing may be identified by the subject of cognition as objects $p_i$ with properties $f_i^p$ and their set $\{p; f^p\}$ as the object $P$. The set $P:\{p; f^p\}$ will be called a ***partition*** of the object $P$ onto the ***elements*** $p_1, p_2,...$ by ***laws*** $f_i^p$. The object $P$ will be called ***generalizing*** the elements of its partition; the object, which does not have the generalizing one, the ***most general***; and objects, which do not have the basic ones, will be called ***inseparable***. The partition consisted from inseparable objects will be called ***fundamental***; and a fundamental partition with ***arranged*** objects will be called a ***space***.

The partitions $P:\{p; f^p\}, Q:\{q; f^q\},...$ will be called ***compatible*** or ***redundant***, if there exists some common ***basic*** partition $E:\{e, f^e\}$, which objects may be identified in these partitions and any objects of compatible partitions may be presented through them. The partitions will be called ***incompatible*** or ***irredundant***, if such common basic partition does not exists. Objects and interconnections of compatible partitions will be called ***normal***, and of incompatible partitions as ***off-site*** to each other or ***strange*** ones.

The notion of a space has a fundamental meaning for perception of the Existing by means of traditional methods of cognition, because exactly in space the subject identifies such basic characteristics of objects as their forms, positions, interconnections. The partitions based on spaces will be called *representation systems*, and as far as the space is a source of initial notions and definitions for representation systems, the irredundant representation systems will be different by *ontology*.

> *Lemma* (*incompatibility and ontology*). Representation systems based on incompatible spaces are incompatible with each other on the base of any space, so they have different ontology.

■ Let the representation systems $P$ and $Q$ based on incompatible spaces $X$ and $Y$ are compatible with each other on the base of some space $Z$. This way the laws of the space $Z$ do not contradict to laws of spaces $X$ and $Y$ and they are compatible with each other on the base of $Z$, which contradicts to initial conditions. Therefore, the space $Z$ does not exist and systems $P$ and $Q$ are incompatible with each other on the base of any space, i.e. initial characteristics and interconnections of their objects do not correlate and objects have different `nature' or ontology. ■

**Example 1** (*simplest partitions*). If objects $e_i$ of the partition $E:\{e, f^e\}$ may be identified in partitions $P:\{p; f^p\}$ and $Q:\{q; f^q\}$ (see Figure 1A), so they will be compatible with each other on the base of the common basic for $P$ and $Q$ partition $E:\{e, f\}$, because objects $p$ and $q$ may be represented through objects $e_i$ and their interconnections $f$: $p_1 = (e_1, e_2, f_{12})$, $p_2 = (e_3, e_4, f_{34})$, $f^p = (f_{13}, f_{14}, f_{23}, f_{24})$ and $q_1 = (e_1, e_3, f_{13})$, $q_2 = (e_2, e_4, f_{24})$, $f^q = (f_{12}, f_{14}, f_{23}, f_{34})$. If objects $e_i$ are also considered inseparable in partitions $P$ and $Q$, so the partition $E$ will be *fundamental* for them. However, if objects $e_i$ cannot be identified in $P$ and $Q$, so these partitions will be *irredundant* with each other.

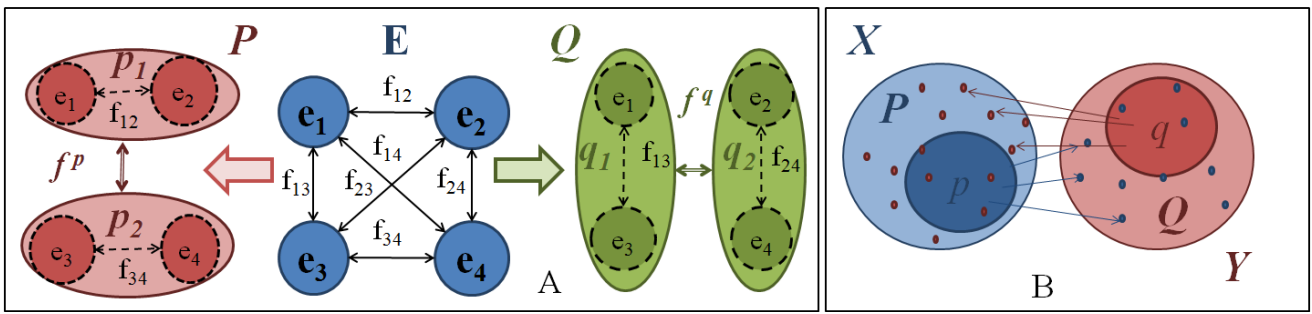

**Figure 1.** Simplest partitions (A) and irredundant representations (B).

**Example 2** (*irredundant representations*). Let some object $P$ is represented as a circle on the plane $X$, which is a space for this object (see Figure 1B). Let's take all points of these circle, mix them carefully and form from them on the plane $X$ the circle $Q$, the same as $P$. Thus, circles $P$ and $Q$ will differ from each other by the point arrangement. Let's form the plane $Y$ by rearranging the points of the plane $X$, so points of the circle $Q$ will be arranged in $Y$ and $Q$ may be identified as an object in the space $Y$. The circle $p$ on the plane $X$ inside the object $P$ will correspond to chaotically distributed points inside the object $Q$, and $p$ cannot be identified as a circle and an object in the plane $Y$. Analogously, the circle $q$ on the plane $Y$ inside the object $Q$ cannot be identified as an object in the space $X$. Thus, the object $Q$ cannot be represented as a partition in the space $X$ and the object $P$ in $Y$, so both objects $p$ and $P$ will be strange in $Y$, while both $q$ and $Q$ in $X$, in spite of $P$ and $Q$ may be identified `as a whole' as objects in both spaces. So as objects $p$, $P$ and $q$, $Q$ have different ontology, the spaces $X$ and $Y$ are off-site to each other.

As far as there is point-to-point correspondence between planes $X$ and $Y$, and objects $P \leftrightarrow Q$ and $p \leftrightarrow q$ consist from `the same' points, one can formally consider them as compatible on frames of the partition of unarranged points of panes. However, this way such basic properties of objects as the shape, the place, interconnections, which cannot exist beside the notion of a space, will be lost, so spaces *determine* the ontology of representations on them.

*Example 3 (compatible spaces).* Let objects $p$ of the representation $P$ are described by functions $p(x)$ in the space $X$, and objects $q$ of the representation $Q$ by functions $q(y)$ in the space $Y$. Let, furthermore, the correspondence: $y = f(x)$, where $f$ is a continuous function having the reverse one: $f^{-1}$, is specified between points of these spaces. In this case any object $q$ of the space $Y$ may be identified as the object $p$ of the space $X$: $q \leftrightarrow q(y) = q[f(x)] = p(x) \leftrightarrow p$, and vice versa: $p \leftrightarrow p(x) = p[f^{-1}(y)] = q(y) \leftrightarrow q$, i.e. spaces $X$ and $Y$ are *basic* for each other and *redundant*, and objects $p$ and $q$ are normal for both spaces.

An example of redundant spaces in physics is *the Minkowski space* in special relativity, which is a set of inertial frames of references. The coordinates of inertial frames are interconnected one-to-one by Lorentz transformations, which keep the arrangement of events in them and in accordance with *the relativity principle* frames are equivalent for the description of the surrounding world, so each frame may be basic for others. Thus, inertial frames of references of the Minkowski space are redundant spaces, and in classical physical theories the subject-observer deals with *normal* objects of inertial frames.

*Example 4 (irredundant spaces).* Let objects $p$ of the representation $P$ are described by functions $p(x)$ in the space $X$, and objects $q$ of the representation $Q$ by functions $q(y)$ in the space $Y$, but, in contrast to the previous example, the one-to-one correspondence is introduced not between spaces, but between objects $p$ and $q$ by the Fourier transformations: $q(y) = \frac{1}{(\sqrt{2\pi})^n} \int p(x) e^{ixy} dX$, $p(x) = \frac{1}{(\sqrt{2\pi})^n} \int q(y) e^{-ixy} dY$, where $n$ is the dimensionality of spaces $X$ and $Y$. In this case there is no one-to-one correspondence between points of spaces $X$ and $Y$: any elementary object $x_0$ of the partition $P$ corresponds to the delta-function $\delta(x - x_0)$ in the space $X$, but in the space $Y$ it corresponds to the function $q(y) = \frac{1}{(\sqrt{2\pi})^n} e^{ix_0 y}$ (when $x_0 = 0$: $q(y) = (\sqrt{2\pi})^{-n}$. Analogously, any elementary object $y_0$ of the representation $Q$ corresponds in the space $Y$ to the delta-function $\delta(y - y_0)$, but in $X$ to the function $p(x) = \frac{1}{(\sqrt{2\pi})^n} e^{-iy_0 x}$ (when $y_0 = 0$: $p(x) = (\sqrt{2\pi})^{-n}$. Therefore, the space $X$ is not basic for objects $q$, which are strange for it, and $Y$ for $p$. Thus, spaces $X$ and $Y$ are *irredundant*; objects $p$ and $q$, in spite of one-to-one correspondence, are off-site to each other, and representations $P$ and $Q$ have different ontology.

Fourier transformations are in foundations of the *quantum-mechanical approach* in physics, so the relativistic and quantum-mechanical approaches correspond to irredundant representations, what explains the impossibility of efforts to unify them in one theory, traditionally operated in frames of one representation system. Thus, quantum-mechanical objects are strange for relativistic theories, and their paradoxical corpuscular-wave properties connected with their different ontology. The further analysis gives the additional confirmations of this supposition.

## 3. Strange objects in physics

Due to connectivity of the Existing, all its parts act to each other. The interaction of irredundant spaces will be called their **coherence**, generally, in some **coherence regions**. The result of the coherence will be called the **coherence object**, which appears as the object in the coherent spaces. For example, objects $P$ and $Q$ on Figure 1B are the result of coherence of irredundant spaces $X$ and $Y$ in the circular coherence region, which is also limited the field of existence of objects $p$ and $q$.

The coherence of spaces $X$ and $Y$ in regions $X_Y \subseteq X$ and $Y_X \subseteq Y$ will be noted as: $X \supseteq X_Y \rightleftharpoons Y_X \subseteq Y$. In general case the formation of the coherence objects of two $\{X_Y \rightleftharpoons Y_X\}$, three $\{X_{YZ} \rightleftharpoons$

$Y_{XZ} \rightleftharpoons Z_{XY}$} and more coherent spaces (see Figure 2A) is possible. If the whole space is the coherence region, this space will be called **embedded** into the **enfolding** one. The space $Y$ on Figure 2B is embedded into the enfolding space $X$: $X \supset X_Y \rightleftharpoons Y_X \equiv Y$; and $X$ is embedded into the enfolding space $Z$: $X \equiv X_Z \rightleftharpoons D_X^Z \subset Z$.

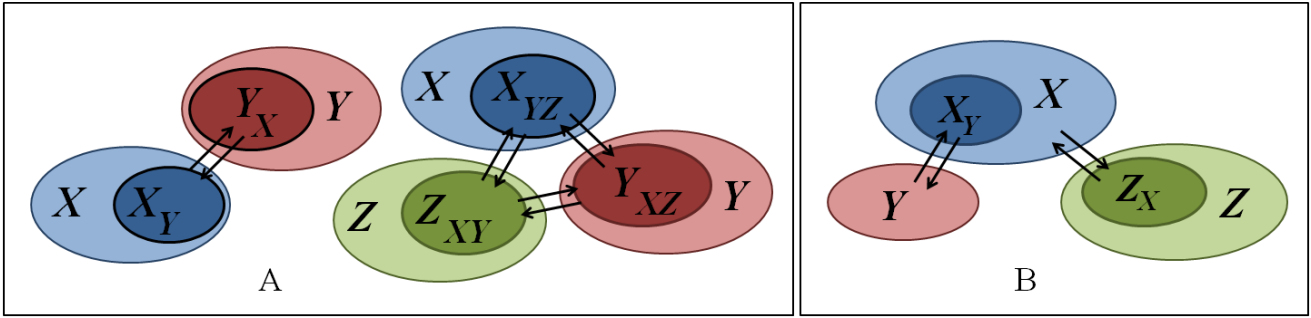

**Figure 2.** The space coherence (A), embedded and enfolding spaces (B).

### 3.1 Relativistic and quantum theories

In classical theories (Newton physics, special and general theories of relativity) the behavior of the physical object or system is described with help of the principle of least action for the integral: $S = \frac{1}{c}\int \Lambda \sqrt{-g}dX$, where $S$ is an action, $c$ a speed of light, $\Lambda$ a Lagrangian density of the system, $dX$ the volume element of the space-time $X$ with the metric tensor $g_{ik}$ and the determinant $g$. In Galilean coordinates: $g = -1$ and $S = \int L dt$, where $t$ is a time, $L = \int \Lambda dV$ a Lagrangian of the physical system [8].

If the physical object is a result of the space coherence in region $D \subset X$, so, separating this region in action $S$, one can get:

$$S = S_f + S_m = \frac{1}{c}\int \Lambda_f \sqrt{-g}dX + \frac{1}{c}\int_D \Lambda_m \sqrt{-g}dX, \tag{5}$$

where $S_m$ and $\Lambda_m$ describe the physical object itself, and $S_f$ and $\Lambda_f$ describe its interconnections (fields) in the space $X$. In classical theories the physical object is described in frames of the space-time continuum, so is considered as a set of material points. This way the function $\Lambda_m$ gets sense of density distribution of the material points in the region $D$. Such representation corresponds to the replacement of the strange object by the normal one, what in fact remove from consideration the specific processes in and around the coherence region. It looks like, for example, to replace the strange object $Q$ from Example 3 by the normal one $P$ with quite different internal structure. Such description cannot be satisfactory in and near the coherence region.

In and near the coherence region one needs to take into account that the physical objects $p,q,r,...$ in the coherent spaces $X,Y,Z,...$ are caused by the coherent object, which state depends on functions $p(x),q(y),r(z),...$, determined in corresponding spaces. If some operators $\hat{f}_{ij}$ described the interactions of strange objects to each other are introduced, so the state of the coherence object may be described by the iteration equations for the vector of functions with the iteration parameter $\xi$:

$$\begin{pmatrix} p_{\xi+1} \\ q_{\xi+1} \end{pmatrix} = \begin{pmatrix} \hat{f}_{11} & \hat{f}_{12} \\ \hat{f}_{21} & \hat{f}_{22} \end{pmatrix}\begin{pmatrix} p_\xi \\ q_\xi \end{pmatrix}, \begin{pmatrix} p_{\xi+1} \\ q_{\xi+1} \\ r_{\xi+1} \end{pmatrix} = \begin{pmatrix} \hat{f}_{11} & \hat{f}_{12} & \hat{f}_{13} \\ \hat{f}_{21} & \hat{f}_{22} & \hat{f}_{23} \\ \hat{f}_{31} & \hat{f}_{32} & \hat{f}_{33} \end{pmatrix}\begin{pmatrix} p_\xi \\ q_\xi \\ r_\xi \end{pmatrix}, \dots . \tag{6}$$

Equations (6) transforms into usual integro-differential equations: $p_{\xi+1} = \hat{f} p_\xi$ analogous to (5), if the influence of the strange objects is not taken into account. This way, the iteration parameter $\xi$ is some space, time coordinates, an interval, etc. in space $X$.

In general case the interaction operators $\hat{f}_{ij}$ are unknown, but the properties of the physical objects $p,q,r,...$ put limitations onto the matrix operator $\hat{F}$. For example, the condition of the coherence object stability may be described by the equation:

$$\hat{F}^n \equiv \hat{I}, \qquad (7)$$

where $\hat{I}$ is an identical operator corresponding to the unit matrix, $n$ is the number of iterations needed to put the system in initial state: $(p_{\xi+n}, q_{\xi+n}, r_{\xi+n}, ...) \equiv (p_\xi, q_\xi, r_\xi, ...)$. For example, in case of the coherence of two spaces, one needs two reflections, i.e. two iterations: from $X$ to $Y$ and back, to put the system into initial state (see Figure 2A), so, according to (7), one can get: $\hat{F}^2 \equiv \hat{I}$. In case of the coherence of three spaces, one needs three reflections, so: $\hat{F}^3 \equiv \hat{I}$. This way, two variants of reflections are possible: $X \to Y \to Z \to X$ and $X \to Z \to Y \to X$, which the observer may interpret as the *spin* of the object.

In *quantum mechanics* the physical object corresponds to the *wave function* $\Psi = \Psi_X + i\Psi_Y$ defined in Hilbert space, which in fact is a state vector of two functions, so corresponds to the coherent object of two spaces. In this case the condition of stability $\hat{F}^2 \equiv \hat{I}$ transforms into condition of unitarity of operator $\hat{U}$ affecting on the wave functions: $\hat{U}^* \cdot \hat{U} \equiv \hat{I}$, where $\hat{U}^*$ is the complex conjunct operator. For Hermitian operator $\hat{H}$ the unitary operator will be: $\hat{U} = \exp(i\hat{H})$. Thus, quantum mechanical approach may well be considered as a particular case of the space coherence (6) with stability condition (7). The particle spin is described by well-known Dirac matrices.

The coherent objects of two and three spaces cannot be represented as the coherent objects of lower number of spaces, so cannot decay and are stable, while the coherent objects of four and more spaces can decay on objects of lower number of spaces, so may be *unstable*. For example, the coherent object of four spaces may decay on two coherent objects of two spaces; the coherent object of five spaces may decay on two coherent objects of two and three spaces, etc. This way, together with usual conservation laws, the spin, as the established order of off-site space interactions, also needs to be conserved. *Resonances, unstable particles and nuclei* may be examples of the *decay* of the coherent objects of multiple spaces. One usually deals with the coherent objects of multiple spaces during *interactions of high energy particles*.

Such observed phenomena in elementary particle physics as paradoxical confinement of protons and neutrons in nuclei (it is a quark confinement in QCD) well correspond to the model of the multiple space coherence, and this way there need no additional suppositions about the existence of new fundamental strong interactions for confinement in the coherent region. The paradoxes appear exactly when one tries to consider the strange objects as normal ones. Some methods of mathematical description and quantization of coherent objects are proposed in [9].

The model of space coherence not only gives a possibility of general description of physical objects of relativistic and quantum theories, but reveals a reason of appearance of the physical objects themselves, which usually are postulated as parts of the `objectively existing material world'.

### 3.2 Cosmological objects

Such observed cosmological phenomena as an anomalous attraction of visible matter in galaxies and the accelerated expansion of the universe also well described by the model of coherent spaces. This way, additional suppositions about the `inherent space curvature' or about the existence of *dark matter* and *dark energy* of unknown nature are not needed

***Example 5*** *(embedded and enfolding spaces)*. Let coordinates of spaces $X$, $Y$ and $Z$ are interconnected as: $x = f(y)$ and $z = f(x)$, where $f$ is increasing, continuously differentiable and limited function (for example: $f(\xi) = \text{arctg}(\xi)$, $f(\xi) = \xi/|\xi| \cdot [1 - \exp(-|\xi|)]$, etc.), i.e. the space $X$ is embedded into the space $Z$ and enfolds the space $Y$: $Y \rightleftharpoons X_Y \subset X \rightleftharpoons Z_X \subset Z$ (see Figure 2B).

Due to properties of the function $f(\xi)$, its derivative (or the gradient $\nabla f$ in the multidimensional case) tends to zero $f'(\xi) \to 0$ with $\xi \to \infty$, and the derivative of the reverse function, on the contrary, tends to infinity: $(f^{-1})'(\eta) \to \infty$ on the border of the coherence region. Thus, $dx/dy = f'(y) = 1/(f^{-1})'(x) \to 0$ on the border of $X_Y$, and $dx/dz = 1/f'(x) \to \infty$ with $x \to \infty$.

As far as the one-to-one correspondence is introduced between spaces in coherent regions, off-site objects may be identified, but their behavior will differ from one of the normal objects, which will look strange and unexplainable for the observer. Indeed, instead of the expected velocity: $v_y = dy(t)/dt$, where $t$ is a time in $X$, of off-site objects of embedded space $Y$ in the space $X$, their visible velocity $v_x = dx(t)/dt$ will tend to zero: $v_x = (f^{-1})'(x)v_y \to 0$ near the borders of the coherent region $X_Y$, as if some invisible forces act to them and confine inside the coherent region. Such effect, unexplainable for normal objects, is observed in cosmology as the anomalous confinement of the visible matter of galaxies, what is `explained' by the gravitational attraction to the invisible ***dark matter*** of unknown nature.

In contrast, the visible velocity $v_x$ of objects of enfolding space $Z$ in $X$ will be observed as increasing with $x \to \infty$: $v_x = v_z/f'(x) \to \infty$, so in the space $X$ these objects will be observed as flying away with the acceleration. Such effect in cosmology is interpreted as the accelerated expansion of the universe.

Formally, the expansion of the universe may be described in the general theory of relativity by adding the well-known cosmological term $\Lambda_0 g_{ik}$ into Einstein equations for gravitational fields. However, the appearance of such term is unexplainable in frames of general relativity, because means "*the assignment to the space-time the inherent curvature connected with nor matter nor gravitational fields*" [8]. The matching of the cosmological term with the energy-momentum tensor $T_{ik}^{DE}$ of some invisible matter – dark energy of unknown nature uniformly distributed over the whole space leads to equality: $\frac{8\pi k}{c^4} T_{ik}^{DE} = \Lambda_0 g_{ik}$, what with $g_{ik} = \text{diag}\{1, -1, -1, -1\}$ corresponds to: $T_{ik}^{DE} = \text{diag}\{\rho, -\rho, -\rho, -\rho\}$, where $\rho = \Lambda_0 c^4/8\pi k$, i.e. with positive energy density of the introduced dark energy, its pressure and momentum will be negative. In distinction with the `unexplainable curvature' of space-time or unknown dark energy with unphysical properties, the behavior of strange objects of the enfolding space has clear physical sense and does not need additional interpretations. Uncertainties, paradoxes and contradictions appear exactly when the strange objects are compared with the normal ones.

### 3.3 The universe formation and the physical vacuum

The freedom principles during the universe formation V. Dokuchaev analyzed in [10] by comparison with possible `ways of existence of albuminous bodies':

- `English': it is allowed everything what is not forbidden;
- `German': it is allowed only that what is not forbidden;
- `French': it is allowed everything and even that what is forbidden;
- `Chinese': it is forbidden everything and even that what is allowed;

and had concluded that the universe formation corresponds to the `English' freedom principle. In contrast to the universe, the Existing includes different ontological systems with mutual contradictions and paradoxes inherent to them, so the freedom principle of its formation, let's call it `Russian', needs to compile all principles considered above, where everything allowed relates mainly to off-site systems, but everything forbidden to own one:

- `Russian': it is allowed everything what is not forbidden and even that what is forbidden, but not in own system, where it is allowed only that what is not forbidden and it is forbidden everything and even that what is allowed.

In this connection, the question: Whether all partitions from possible ones may exist and if not, so which ones really exist? – assumes a great importance. From the point of view of the declared freedom principle a supposition about the existence of any ones looks the most consequent. In this case the observer can perceive only those, which coherence with his own space he may identify. Other partitions the observer may `perceive' as the influence of something uncertain, that he cannot identify as normal objects. Mainly, these influences will be chaotic and multidirectional, so compensate each other as, for example, the virtual particles of the *physical vacuum*. If the influences accidentally or not are correlated with each other, they may be well identified. It may seem, for example, as the *appearance* of particles from the physical vacuum near the black hole gravitational radius or as the *birth* of the universe in a result of the physical vacuum accidental fluctuation.

External zones of coherent regions of enfolding space (the zone $Z \setminus Z_X$ on Figure 2B) look unreachable for cognizers from embedded spaces, what on the first sight contradicts to connectivity of the Existing. However, this contradiction is visible and has quite clear explanation: all parts of the Existing are interconnected, but connections with external regions of enfolding spaces are unreachable for representation systems of embedded spaces, because are out of frames determined by their ontology.

The inaccessibility of representations outside the own system puts face-to-face with the problem of understanding in cognition.

## 4. The problem of understanding in science

Traditionally, the problem of understanding is closely connected with the availability of *consistent* representation system, i.e. the perfect theory in science, which may describe all observed phenomena of the surrounding world. This way, everything, that is not or out of the frames of the system, is perceived as beyond the ***understanding***, so the Einstein's tendency `to put into system' corresponds to the wish to understand `all cognizable phenomena' in the surrounding world. However, the founder of the relativistic picture of the world was not able to do it. The stumbling block was the arising at that time the quantum mechanics with its strange dual objects, uncertainties and paradoxes. "*God does not play dice*" – said A. Einstein, pointing out figuratively that not only the notions and the apparatus of the quantum mechanics are incompatible with the approach of the relativistic theories, but they violate the basic principles of the relativistic picture of the world. But he has got not less figurative answer from N. Bohr: "*Do not order God what to do*". In any way, many efforts `to put in the system' the relativistic and quantum theories were unsuccessful. The quantum mechanics is still "*full of mysteries and paradoxes discipline, which we do not sufficiently understand, but are able to use*" (M. Gell-Mann [11]), a discipline, which "*nobody understands*" (R. Feynman [5]).

The not so far `discovered' dark matter and dark energy in cosmology are also no less strange. Their nature is unknown, and they were introduced to ground the anomalous attraction of visual matter in galaxies and the accelerated expansion of the universe correspondingly. It looks impressive that, according to estimations, the dark matter and dark energy of unknown nature are more than 95 % of substance in the universe that characterizes the amount of unknown, probably, not final.

It is not all problems of the understanding in science: the system approach meets principal objections from the system approach itself. In the mathematical set theory it is expressed in system paradoxes of Cantor and Russell [12-15]. *The Cantor paradox* puts in doubt the existence of the most general set of all sets (i.e. the one consistent and universal representation system), the cardinality of such set will always be lower than the cardinality of its power set containing all subsets. *The Russell antimony* calls in question the general possibility of classification and systematization with help of sets (and representations). Indeed, if to define the set, which is not containing itself as an element, as `normal' and a set, which is containing itself as an element, as `abnormal', so the Russell set $R$, i.e. the set of normal sets cannot be classified nor as normal, nor as abnormal. Really, if $R$ is supposed as normal, it is abnormal, because contains abnormal sets, but if it is supposed as abnormal, it is normal, so as contains normal sets. In both cases we come to contradiction, so the Russell set cannot be classified.

There is no the acknowledged way to overcome the system paradoxes of the set theory. The existing ways to eliminate paradoxes by putting some limitations on sets and operations with them, from the physics point of view looks like another efforts `to order God what to do'. Besides of that, the uncertainties and paradoxes appear in properties of observed physical objects, so cannot be eliminated from the most general system. A science comes closely to principal contradiction: the system approach needs to eliminate uncertainties and paradoxes, but they are inherent parts of the surrounding world, so one needs to eliminate that what cannot be eliminated. Exactly this contradiction was pointed out in Introduction as a crisis of system approach in scientific cognition.

A lot of unsuccessful efforts to unify the relativistic and quantum theories, the `discovery' of cosmological objects with unphysical properties, the unsolvable system paradoxes have leaded to the introducing of the new scientific paradigm: the mathematical formalism. The apologists of this paradigm assert that the science intrudes into regions, which are so far from everyday representations, that it is impossible to find correlations of phenomena being discovered with the existing representations, so one needs to describe these phenomena mathematically with help of equations without understanding the sense of introduced notions or simply without introducing any notions at all. This way, the apologists of the mathematical formalism answer to opinions like `nobody understands the quantum mechanics' by the Dirac's words: "*There is nothing to understand!*" and also cite to the words of Galileo Galilei, who asserted centuries ago that "*Nature is written in the language of Mathematics, and we must learn this language*".

Such arguments are absurd, because the Galilei's words were appealed to his opponents, who considered that the heliocentric model, even corroborated by the Galilei's observations and calculations, does not need at all, because one can well determine the planet position on the sky with help of existing tables, and it does not matter what is rotating about, moreover, it is doubtful whether t will be possibility to test it. Thus, Galilei supported exactly the understanding of the laws of nature, in particular, the laws of the planet motions, but his opponents preferred to use formally the tables. The present-day formalists also prefer to use formally the mathematical equations without the understanding and, at that, covering themselves by the Galilei's words...

There arises the paradoxical and dangerous situation when the science in the face of the mathematical formalism apologists are denying from its main advantages – the systematization and the understanding, and this tendency are becoming the dominant *paradigm* impeding the science development. According to the Tomas Kuhn's theory [16], it is quite difficult to change the dominant paradigm because of the counteraction of its apologists, so the changing or replacing the paradigm often occurs by means of the scientific revolutions. The T. Kuhn's theory was criticized by a number of scientists for `the misunderstanding of the principal meaning of the correlation principle between old and new theories, for the lack of real historicism, for misunderstanding of the heterogeneity of science development', etc. (V.L. Ginzburg [17]). However, later, in full correspondence with the statements of Kuhn's theory some critics created the so-called 'Commission for struggle (!) with pseudo-science'. It is not difficult to guess that for the dominant paradigm apologists the pseudo-science is everything that is not in frames of the paradigm. The Galilei's opinions were also declared as heretical by whose, who considered themselves as the only preservers of Truth and Knowledge, and the science was thrown back by the centuries…

But what the system approach and the understanding may be talked about in the concept of incompatible representations, if such representations cannot principally be unified in frames of one system, and exactly the impossibility to create the unified theory has lead to denying the understanding and was the reason for introduction of the mathematical formalism? Moreover, this concept does not propose a way to solve the principal system paradoxes of the set theory, but only confirm the impossibility to create the most general system.

It is right, but, nevertheless, the concept of incompatible representations gives a possibility to solve all of these problems by some unexpected specific way. First of all, separating the irredundant systems, the concept removes from them the unavoidable ontological contradictions, so keeps their internal consistency and the understanding in frames of each of these systems. This way, the irredundancy of systems is a necessary condition, because otherwise, the unavoidable ontological contradictions could be eliminated. Thus, the concept gives a principal possibility to solve the problem of understanding in science and to overcome the crisis of the system approach in scientific cognition.

Secondly, the existence of many irredundant systems means the existence of different laws and logics in them, so one can well use *different* methods of eliminating the system paradoxes, also ones from the set theory (see [18] for more details) without `the pointing out God what to do', because any system or logic does not deny the existence of others, even incompatible ones.

In this connection, the theoretical investigations of interactions of irredundant systems and incompatible logics become quite important. Being the parts of one and connected Existing, the irredundant representation systems act and form each other, defining the ***evolution*** of the Existing. The freedom principle does not limit the type of irredundant systems, but these types determine the result of their interactions. For example, correlated space coherence may lead to the creation of physical particles or the universe, but uncorrelated one to formation of virtual particles of physical vacuum.

In fact, the concept of incompatible representations is exactly the unified theory containing the relativistic and quantum theories, the strange cosmological objects, but this unification occurs on the basis of the generalized method of cognition outside of the ontological frames of representation systems, where the notions of objects and even systems are disappeared. From the traditional method of cognition point of view, the unification occurs by means of the separation of ontologically incompatible systems.

Thus, the concept of incompatible representations expands considerably the possibilities of the scientific cognition by spreading it onto many different ontological representation systems, by eliminating the principal paradoxes in the system approach of the scientific cognition and unifying many incompatible approaches into the multi-polar scientific picture of the world.

## 5. Multi-polar picture of the world

The generalized approach to cognition by definition has to be on the background of any human representations from everyday and religious ones up to scientific theories, philosophical systems and the worldview. The Existing contains different systems including the incompatible ones, so contains uncertainties, paradoxes and contradictions between them.

For example, the ***material world***, determined in materialistic approach as some `objective reality given in senses', influences a man at least through his senses and, by definition, is a part of the Existing. The ***subjective representations***, understood as images of the material objects, differ from them `by nature', but influence a man by determining his behavior and are also parts of the Existing.

Representations, notions and ideas, accepted in the society, i.e. ***common notions***, such as the language, laws, moral rules, scientific and religious views, etc., also influence a man and are parts of the Existing. But they differ both from material objects, because are representations, and from the subjective representations, because do not directly relate to the person, so are objective to him. Inherently, the common notions are the product or the sublimation of subjective representations of *many* persons, what defines the objective character for each of them.

In materialistic approach the relative objectivity of common notions differs from the absolute objectivity of the material world, which is completely independent from the cognizing person. However, the cognizers cannot identify strange objects and the cognizers with different ontology, so their influences will be perceived as something having the absolutely objective character. One may suppose that also as a sublimation of subjective representations into common notions, the sublimation of common notions with different ontology will form the absolutely objective ***ideal world***, which is an initial point in *objective idealism*.

Thus, the *concept of irredundant representations* reveals the ontological essence of the separation of such philosophical schools as materialism, subjective and objective idealism and interconnects them together as interacting, supplementing and forming each other irredundant representation systems. Indeed, the *material world* surrounding us forms the *subjective representations*, from which the *common notions* are sublimated, also as there is formed from different ontological notions the *ideal world*, which is a source of *objective physical laws* and *objects* in each of surrounding worlds with different ontology, in particular, in our *material world*. Such ***cyclic chains*** of irredundant essences interacting and forming each other determine the ***evolution*** of the Existing, and the contradictions between irredundant essences appear as the moving forces of evolution.

From this point of view it becomes obvious the boundedness of the traditional for *natural sciences* materialistic approach in cognition, mainly operated in frames of one ontological system, from which the dialectical unity of irredundant systems is perceived as paradoxical, full of uncertainties and contradictions. The concept of irredundant representations gives a possibility to come out from the frames of materialistic approach and solve many principle problems of systematic approach in scientific knowledge, at the same time interconnecting many scientific theories and approaches. For example, the conclusion made before that the ontology of the subject of cognition defines the ontology of his representation system and the world surrounding him may be in foundations of the anthropic principle [19,20] in cosmology. And the fractal reflections of B. Mandelbrot [21,22], which are in foundations of I. Progogine's theory of dissipative structures [23,24], may be considered as the particular case of iteration reflections of coherent spaces.

Confining itself by the study the material world only and eliminating the contradictions from the corresponding representation system, the materialistic approach removes from consideration the moving forces of evolution of the Existing and a material world in particular. In this connection it looks quite naïve to explain the birth of the material world while staying in frames of the materialistic approach. The concept of irredundant representations reveals the moving forces of the evolution, which are necessary for foundation of the principle of self-organization of matter in synergetic theories, such as a theory of self-organization of the universe of V. Branskiy [25].

The scientific approach has reasons to criticize the religious systems for their inconsistency and groundlessness, but exactly these imperfections give them a possibility to go out the frames of the materialistic approach and come closer to perception of the Existing in dynamics, together with contradictions and paradoxes inherent to it. For example, in **Hinduism** the trinity of supreme divinity – the Trimūrti (in transcription from Sanskrit: "three forms") unifies a triad of deities, typically personified as Brahma the creator, Vishnu the preserver, and Shiva the destroyer or regenerator, which corresponds to the cosmic functions of creation, maintenance and destruction. These counteracting forces are incompatible with each other, but exactly the contradictions between them appear as the moving forces of the evolution of the Existing. Moreover, the self-organization and evolution is possible only in their unity, and only their irredundant essence does not let them to destroy each other. Such forces of creation, maintenance and destruction are in the basis of **Buddhist circuit of Saṃsāra**, corresponding to cyclic chains of evolution considered before, which are duplicated on different levels and parts of the Existing. In **Christianity** the unity of the Existing is represented not as the unity of counteracting forces, but as the unity of irredundant forms or essences of different nature. Indeed, the basic notion of the Saint Trinity are personified as God the Father, God the Son and the Holy Spirit, which have clear analogies with ideal (the absolute ideal world), material (the material incarnation of God the Father) and subjective (moving force and the basis of formation) essences correspondingly. The interaction of such essences were described before in analysis of philosophical schools. Analogies with irredundant representations give a possibility to interpret the unity of God not as *monotheism*, which implies one system, the undivided authority and *uniformity* of the world, but as a unity of the world in the *diversity* of a lot of interacting and forming each other irredundant essences in it, what is principally *opposite* the uniformity.

In a *social* plane, a unity in diversity corresponds to the concept of the *multi-polar* world understood as a unity of closely interconnected, interacting and forming each other *different* social systems. Widespread, but principal mistake is the understanding of the multi-polar world as an aggregate of identical `poles', i.e. the social systems of the same type, what, as it was shown in [26], causes the competitive struggle and is one of the stages of globalization on the way to uniformity and undivided authority opposite to diversity. The unity in diversity means the close interconnection of principally incompatible social systems, which eliminates their merging, such, for example, as monarchism, capitalism and socialism. The interconnection of irredundant social systems defines their mutual development and evolution, but efforts of their mechanical merging by artificial mixing of people and by introduction of `tolerance', in contrary, will lead only to the worsening of insuperable contradictions, to conflicts and general degradation of forcibly `unifying' social systems, which is exactly observed in different regions of the modern world, where the globalization is trying to be introduced.

The one, closed and connected Existing unifies many representation systems with different ontology, i.e. many perceived surrounding worlds irredundant with each other. The ontological incompatibility and irredundancy of systems and worlds does not mean their isolation from each other. On the contrary, being parts of one Existing, they are closely interconnected, mutually supplement, enrich and improve each other without mixing and staying different `by nature'. Such cyclic chains of irredundant essences, generated on different levels, determine the evolution of the Existing, and the contradictions between irredundant essences appear as the moving force of the evolution.

### Conclusion

The introduced postulate of existence and cognoscibility allows to specify many basic notions of the theory of knowledge, to separate unknown and incognizable, to generalize the process of cognition of the Existing onto any interactions of its parts, what reveals the boundedness of the traditional methods of cognition, consisting in an identification of objects of the surrounding world, by the nature, i.e. the ontology of the cognizing subject. Thus, the generalized process of cognition leads to the concept of irredundant representations, i.e. to the possibility of existence of different by ontology representation systems incompatible with each other on the level of basic notions and definitions, so irredundant on frames of one consistent system.

By separating the incompatible representation systems, the concept of irredundant representations eliminates the internal contradictions in each of the systems, so solves the problem of the system approach in scientific cognition and also solves the interconnected problem of understanding in science. In fact, the concept of incompatible representations is the Unified theory containing the relativistic and quantum theories, the strange cosmological objects, but this unification occurs on the basis of the generalized method of cognition, so outside the ontological frames of representation systems, where the notions of objects and even systems are disappeared. From the traditional method of cognition point of view, the unification occurs by means of the separation of ontologically incompatible systems.

The one, closed and connected Existing unifies many irredundant representation systems with different ontology, corresponding to many surrounding worlds. Being parts of one Existing, ontologically incompatible and irredundant representation systems and worlds are closely interconnected, mutually supplement, enrich and improve each other without mixing and staying different by nature. Arising cyclic chains of interconnected irredundant essences, generated on different levels of the Existing, determine its evolution, and the contradictions between irredundant essences are the moving forces of the evolution.

The concept of incompatible representations expands considerably the possibilities of the scientific cognition by spreading it onto many ontologically different representation systems, this way, eliminating the principal paradoxes in the system approach of scientific cognition and unifying many incompatible approaches. The generalized method of cognition by definition are in the basis of different representation systems: from everyday and religious ones up to scientific theories, philosophical systems and the worldview, what gives a possibility of general analysis and specification of their foundations, leads to the multi-polar picture of the world.

## Acknowledgements

I would like to acknowledge Prof. Lewis Ryder, who in many years was my consequent opponent in scientific discussions, in which, as it is known, the truth is born.

А.В. Новиков-Бородин

# ОНТОЛОГИЧЕСКИЕ СИСТЕМЫ В ПОЗНАНИИ

**Аннотация.** Исследуются обобщённые методы познания Сущего, как всего, что может оказывать воздействие на познающего, а всё, отличное от Сущего, постулируется как неотличимое от несуществующего и непознаваемое. Традиционные методы познания, заключающиеся в идентификации объектов окружающего мира, учитывают не все воздействия, так как ограничены природой, то есть онтологией субъекта познания. Онтология определяется на уровне исходных понятий и определений, поэтому даже идентифицированные объекты из систем с различной онтологией выглядят неопределёнными и парадоксальными, точно такими как, например, объекты квантовой механики или тёмная материя и энергия в космологии. Системы с различной онтологией несовместимы друг с другом на уровне исходных понятий, и не могут быть сведены в рамках одной непротиворечивой системы или теории. Концепция несводимых представлений объединяет на уровне, превышающем онтологические рамки, различные по онтологии системы и несовместимые подходы, такие как релятивистский и квантовый. Концепция позволяет снять принципиальные противоречия системного подхода в научном познании, чем решает проблему понимания в науке. Обобщённые методы познания по определению лежат в основе различных систем представлений: научных, философских, религиозных и социальных, что даёт возможность общего анализа и уточнения их основ. Взаимосвязи некоторых физических теорий, философских, религиозных и социальных систем кратко анализируются с этой точки зрения. Кратко рассматривается многополярная картина мира, к которой приводит концепция несводимых представлений.

**Ключевые слова:** обобщённый процесс познания, странные объекты в физике, несовместимые онтологические системы представлений, многополярная картина мира

## Введение

Наука в стремлении *"свести вместе посредством системы все познаваемые явления нашего мира"* (А. Эйнштейн) всё глубже проникает внутрь материи, всё дальше заглядывает в просторы Вселенной. Её достижения трудно переоценить, но некоторые свойства и закономерности, открытые экспериментально и даже успешно используемые, никак не укладываются в единую непротиворечивую систему представлений – научную картину мира и разрушают *"целостный образ предмета научного исследования"* ([1]: научная картина мира): то пространство и время оказываются относительными; то частицы ведут себя как волны [2-5]; то вещество галактик удерживается невидимой силой неизвестной природы, а сами галактики, наоборот, разлетаются с ускорением [6,7]. При этом выясняется, что невидимые и странные тёмная материя и тёмная энергия неизвестной природы, отвечающие за это удержание и расширение, составляют более 95% вещества во Вселенной, и все предыдущие века физика изучала лишь малую часть 'нормальной' материи.

После многократных неудачных попыток свести подобные наблюдаемые явления в единую непротиворечивую систему, например, путём объединения релятивистского и квантово-физического подходов, установилось мнение, что открытые объекты и закономерности настолько далеко выходят за рамки 'житейского опыта', что попытки понять их, сведя в общую систему представлений и сопоставив с 'обыденными' образами и понятиями невозможны. Подобная парадигма отрицает **понимание** в физике и приводит к математическому формализму, при котором исчезает возможность критического анализа вводимых понятий, возрастает вероятность спекуляций в науке, превращения её в фарс. Дело дошло до того, что *"В современной физике не сохранилось ни одного из классических определений материи. Однако как философия, так и физика предпочитают обходить это ставшее неопределенным и темным понятие, заменяя его другими – пространство-время, хаос, система и др."* ([1]: материя), то есть стало пропадать само понятие предмета и цели научного исследования. Пытаясь избежать рассмотрения подобных вопросов и считая их уделом философии, современная наука отходит от своих основных преимуществ – системного подхода и понимания, без которых её дальнейшее развитие становится проблематичным.

Возможно, системный подход в науке исчерпал себя? Диалектика, например, как *"логическая форма и всеобщий способ рефлексивного теоретического мышления, имеющего своим предметом противоречия его мыслимого содержания"* ([1]: диалектика) считает противоречия неотъемлемой частью познания, а они несовместимы с непротиворечивым системным подходом, лежащим в основе научных теорий и методов познания. Ситуация кажется безвыходной. Впрочем, наука не впервые сталкивается с проблемами, кажущимися непреодолимыми, а в результате оказывается, что непреодолимыми были лишь наши заблуждения, и мы в очередной раз находимся на этапе переосмысления основ и перехода к новым возможностям.

## 1. Обобщённый процесс познания

В *традиционном* для науки материалистическом понимании процесса познания **субъект** познания *идентифицирует* **объекты** *окружающего его мира* через их **свойства**.

> ***Процесс познания*** (*традиционный*) – это процесс идентификации субъектом познания объектов окружающего его мира и их свойств.

Совокупность идентифицированных объектов составляет **представления** субъекта о воспринимаемом им окружающем мире, или **систему представлений**, если эти представления систематизированы. Построение *непротиворечивой* системы представлений традиционно считается целью *научного познания*.

Подобный подход к познанию приводит к целому ряду фундаментальных проблем, связанных неполнотой образов идентифицируемых объектов, с несоответствием их 'реальным объектам', с ограниченностью методов познания и, как следствие, с познаваемостью мира вообще. Не вдаваясь преждевременно в дискуссии по этим вопросам, при общем анализе процесса познания будем исходить из следующего *утверждения*:

> ***Постулат*** (*существования и познаваемости*). Существует для познающего и может быть им познано то и только то, что, так или иначе, прямо или косвенно, оказывает на него какие-либо воздействия.

Строго говоря, то, что '*Существует для познающего*' отличается от '*Всего, что вообще может быть*', ведь вполне может существовать то, что никаких воздействий на познающего не оказывает, но оно будет '*неотличимым от несуществующего*' для него и, следовательно, познано им быть не может. Таким образом, '**Всё**, что вообще может быть' ($\mathcal{W}$) для некого познающего $p$ состоит из 'всего, что может оказывать на него хоть какое-то воздействие' – **существующего** для него или его **Сущего** $\mathcal{E}_p$ и 'того, что никакого воздействия на него не оказывает', то есть логического отрицания Сущего – **Несуществующего** $\mathcal{N}_p = (\neg \mathcal{E}_p)$:

$$\mathcal{W} = \mathcal{E}_p \lor (\neg \mathcal{E}_p) = \mathcal{E}_p \lor \mathcal{N}_p. \qquad (1)$$

Познающий является неотъемлемой частью своего процесса познания, следовательно, *существует* для себя и по определению сам является *частью* своего Сущего, то есть возможность *разделения* Сущего на **части** изначально заложена в самом процессе познания, выделяющем познающего как часть Сущего:

$$p \subset \mathcal{E}_p. \qquad (2)$$

Являясь частью своего Сущего, познающий по определению должен оказывать воздействие на себя, при этом воздействуя и на ту часть Сущего, которой сам является. Таким образом, познающий **взаимодействует** с Сущим *через себя* и **взаимосвязан** с ним через взаимодействия. Взаимодействие Сущего с его частью в *общем виде* определяет **процесс познания**:

> ***Процесс познания*** (*обобщённый*) – процесс взаимодействия Сущего с его частью.

В отличие от познающего в обобщённом процессе познания субъект в традиционном процессе не просто взаимодействует с частями Сущего, но *идентифицирует* их в качестве **объектов**. Обычно идентификация ассоциируется с сознательной деятельностью субъекта, что вызывает дополнительные трудности, связанные с тем, кого можно считать субъектом познания. В случае с познающим подобных проблем не возникает, так как взаимодействовать с Сущим может любая его часть, и *в обобщённом смысле* любую часть Сущего можно рассмотреть в качестве познающего, а изменения, которые часть Сущего претерпевает при взаимодействии с Сущим, вполне можно интерпретировать как ***результат*** познания Сущего его частью. При таком подходе, даже неодушевлённые предметы можно рассмотреть в качестве познающих. Например, улики в следственных мероприятиях или археологические артефакты вполне можно считать свидетелями прошедших событий, а получаемую нами зрительную или звуковую информацию рассматривать как 'показания' электромагнитных полей и звуковых волн. Таким образом, традиционный процесс познания является частным случаем обобщённого.

> ***Теорема*** (*единство Сущего*). Сущее связно и замкнуто для всех своих частей, является единым и единственным для всех познающих в нём.

■ Так как познающий через взаимодействия прямо или косвенно взаимосвязан с любой частью Сущего, то по крайней мере через него все части Сущего взаимосвязаны между собой, и Сущее *связно*. Всё, что хоть как-то связано с Сущим, будет прямо или косвенно связано с познающим в нём и по определению является частью его Сущего. Следовательно, никаких взаимосвязей у частей Сущего помимо взаимосвязей с другими его частями, быть не может и Сущее *замкнуто* для всех своих частей, включая познающих, являющихся его частями.

Пусть познающие $p, q, ...$, имеющие Сущие $\mathcal{E}_p, \mathcal{E}_q, ...$: $p \subset \mathcal{E}_p$, $q \subset \mathcal{E}_q, ...$, являются частями Сущего $\mathcal{E}$: $p, q, ... \subset \mathcal{E}$. Вследствие связности Сущего $\mathcal{E}$, познающие $p, q, ...$ взаимосвязаны, и их Сущие $\mathcal{E}_p, \mathcal{E}_q, ...$ тоже взаимосвязаны через них и являются частью общего Сущего $\mathcal{E}$: $\mathcal{E}_p, \mathcal{E}_q, ... \subseteq \mathcal{E}$. С другой стороны, Сущее $\mathcal{E}$ взаимосвязано с познающими $p, q, ...$ и должно быть частью Сущего каждого из них: $\mathcal{E} \subseteq \mathcal{E}_p, \mathcal{E}_q, ...$. Из условий $\mathcal{E} \subseteq \mathcal{E}_p, \mathcal{E}_q, ... \subseteq \mathcal{E}$ следует:

$$\mathcal{E}_p \equiv \mathcal{E}_q \equiv \cdots \equiv \mathcal{E}, \forall p, q, ... \subset \mathcal{E}, \qquad (3)$$

то есть Сущее *едино* и *единственно* для всех познающих в нём. ■

Таким образом, для любых познающих $p, q, ...$, всё, отличное от их общего Сущего $\mathcal{E}$, неотличимо от несуществующего: $\mathcal{N} = (\neg \mathcal{E})$, и выражение (1) с учётом (3) можно записать в виде:

$$\mathcal{W} = \mathcal{E} \vee (\neg \mathcal{E}) = \mathcal{E} \vee \mathcal{N}, \qquad (4)$$

В дальнейшем, если иное не оговорено особо, будем рассматривать только Сущее $\mathcal{E}$ и познающих только в нём.

## 2. Несводимые представления

Пусть некоторые части Сущего могут быть идентифицированы субъектом как объекты $p_i$ со свойствами $f_i^p$, а их совокупность $\{p; f^p\}$ как объект $P$. Совокупность $P: \{p; f^p\}$ будем называть **разбиением** объекта $P$ на **элементы** $p_i$ по **законам** $f_i^p$. Объект $P$ будем называть **обобщающим** элементы своего разбиения; объект, не имеющий обобщающего, – **наиболее общим**; а объекты, не состоящие из элементов – **неделимыми**. Разбиение, состоящее из неделимых объектов, будем называть *фундаментальным*; а фундаментальное разбиение, объекты которого *упорядочены*, будем называть *пространством*.

Разбиения $P:\{p; f^p\}$, $Q:\{q; f^q\}$,... будем называть *совместимыми* или *сводимыми*, если существует некоторое общее *базовое* разбиение $E:\{e; f^e\}$, объекты которого могут быть идентифицированы в этих разбиениях и через них могут быть представлены любые объекты сводимых разбиений. Разбиения будем называть *несовместимыми* или *несводимыми*, если общего базового разбиения для них не существует. Объекты и взаимосвязи сводимых разбиений будем называть *нормальными*, а несводимых – *сторонними* друг другу или *странными*.

Понятие пространства имеет фундаментальное значение для восприятия Сущего традиционными методами познания, поскольку именно в пространстве субъект идентифицирует такие базовые свойства объектов как их форма, положение, взаимосвязи. Разбиения на основе пространств будем называть *системами представлений*, и поскольку пространство является источником начальных понятий и определений для систем представлений, несводимые системы представлений будут различными по *онтологии*.

> *Лемма* (*несводимость и онтология*). Системы представлений на основе несводимых пространств несводимы друг с другом на основе любого другого пространства и имеют различную онтологию.

■ Пусть системы представлений $P$ и $Q$ на основе несводимых пространств $X$ и $Y$ сводимы друг с другом на базе некоторого пространства $Z$. Тогда законы пространства $Z$ не противоречат законам пространств $X$ и $Y$ и они сводимы друг с другом на основе этого пространства, что противоречит начальным условиям. Следовательно, пространства $Z$ не существует и системы $P$ и $Q$ несводимы друг с другом на основе любого пространства, то есть изначальные характеристики и взаимосвязи их объектов не совпадают и объекты имеют различную 'природу' или онтологию. ■

*Пример 1* (*простейшие разбиения*). Если объекты $e_i$ могут быть идентифицированы в разбиениях $P:\{p; f^p\}$ и $Q:\{q; f^q\}$ (см. Рисунок 1А), то они будут *сводимы* друг с другом на основе общего *базового* для $P$ и $Q$ разбиения $E:\{e, f\}$, так как объекты $p$ и $q$ могут быть выражены через объекты $e_i$ и их взаимосвязи f: $p_1 = (e_1, e_2, f_{12})$, $p_2 = (e_3, e_4, f_{34})$, $f^p = (f_{13}, f_{14}, f_{23}, f_{24})$ и $q_1 = (e_1, e_3, f_{13})$, $q_2 = (e_2, e_4, f_{24})$, $f^q = (f_{12}, f_{14}, f_{23}, f_{34})$. Если при этом объекты $e_i$ считаются неделимыми в разбиениях $P$ и $Q$, то разбиение $E$ будет *фундаментальным* для них. Однако если объекты $e_i$ не могут быть идентифицированы в $P$ и $Q$, то эти разбиения будут *несводимы* друг с другом.

*Пример 2* (*несводимые представления*). Пусть некоторый объект представлен кругом $P$ на плоскости $X$, являющейся пространством объекта (см. Рисунок 1Б). Возьмём все точки этого круга, тщательно их перемешаем и выложим из них на плоскости $X$ круг $Q$, такой же как $P$. Таким образом, круги $P$ и $Q$ будут отличаться друг от друга упорядоченностью точек. Перенумеровывая точки плоскости $X$, образуем плоскость $Y$, в которой точки круга $Q$ упорядочены и он может быть идентифицирован как объект в пространстве $Y$. Круг $p$, определённый на плоскости $X$ внутри круга $P$, на плоскости $Y$ будет соответствовать хаотично разбросанным точкам внутри круга $Q$, и не может быть идентифицирован как круг и объект в пространстве $Y$. Аналогично, круг $q$ на плоскости $Y$ внутри объекта $Q$ не может быть идентифицирован как объект в пространстве $X$. Таким образом, объект $Q$ не может быть представлен в виде разбиения в пространстве $X$, а объект $P$ – в пространстве $Y$, и оба объекта $p$ и $P$ будут странными в $Y$, а $q$ и $Q$ в $X$, несмотря на то, что объекты $P$ и $Q$ могут быть идентифицированы 'в целом' как круг в обоих пространствах. Так как объекты $p$, $P$ и $q$, $Q$ имеют разную онтологию, пространства $X$ и $Y$ являются сторонними друг другу.

Поскольку между точками плоскостей $X$ и $Y$ установлено взаимно однозначное соответствие, а объекты $P \leftrightarrow Q$ и $p \leftrightarrow q$ состоят из 'одних и тех же' точек, формально их можно считать сводимыми в рамках разбиения, представляющего собой неупорядоченную совокупность точек плоскостей. Однако при этом, такие основополагающие свойства, как форма, расположение, и др., которые вне понятия пространства не существуют, будут утеряны, то есть пространства *определяют* онтологию представлений на их основе.

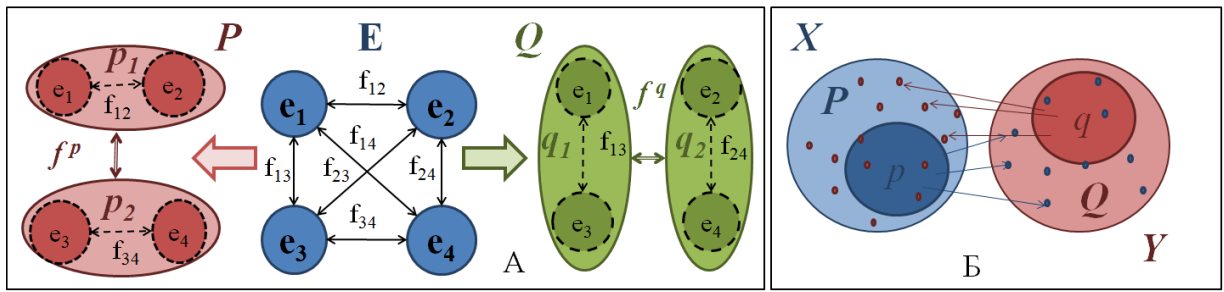

**Рисунок 2.** Простейшие разбиения (А) и несводимые представления (Б).

***Пример 3*** *(сводимые пространства).* Пусть объекты $p$ представления $P$ описываются функциями $p(x)$ в пространстве $X$, а объекты $q$ представления $Q$ функциями $q(y)$ в пространстве $Y$. Пусть, далее, между точками пространств задано соответствие: $y = f(x)$, где $f$ – непрерывная функция, имеющая обратную $f^{-1}$. В этом случае любой объект $q$ пространства $Y$ может быть идентифицирован как объект $p$ пространства $X$: $q \leftrightarrow q(y) = q[f(x)] = p(x) \leftrightarrow p$, и наоборот: $p \leftrightarrow p(x) = p[f^{-1}(y)] = q(y) \leftrightarrow q$, то есть пространства $X$ и $Y$ могут выступать в качестве базовых друг для друга и являются *сводимыми*, а объекты $p$ и $q$ – *нормальными* для обоих пространств.

Примером сводимых пространств в физике является *пространство Минковского*, которое в специальной теории относительности представляет собой совокупность инерциальных систем отсчёта. Координаты инерциальных систем взаимно однозначно связаны преобразованиями Лоренца, сохраняющими упорядоченность событий в них, и согласно *принципу относительности*, эти системы равноправны для описания окружающего мира, то есть каждая из них может быть базовой для других. Таким образом, инерциальные системы являются *сводимыми* пространствами, и в классических физических теориях субъект-наблюдатель имеет дело с *нормальными* объектами инерциальных систем отсчёта.

***Пример 4*** *(несводимые пространства).* Пусть объекты $p$ представления $P$ описываются функциями $p(x)$ в пространстве $X$, а объекты $q$ разбиения $Q$ функциями $q(y)$ в пространстве $Y$, но, в отличие от примера 3, с помощью интегрального преобразования Фурье: $q(y) = \frac{1}{(\sqrt{2\pi})^n} \int p(x) e^{ixy} dX$, $p(x) = \frac{1}{(\sqrt{2\pi})^n} \int q(y) e^{-ixy} dY$, где $n$ – размерность пространств $X$ и $Y$, установлено взаимно однозначное соответствие не между точками пространств, а между объектами $p$ и $q$. При этом, между точками пространств $X$ и $Y$ взаимно однозначного соответствия нет: произвольному элементарному объекту $x_0$ разбиения $P$ в пространстве $X$ соответствует дельта-функция $\delta(x - x_0)$, а в пространстве $Y$ – функция $q(y) = \frac{1}{(\sqrt{2\pi})^n} e^{ix_0 y}$ (при $x_0 = 0$  $q(y) = \frac{1}{(\sqrt{2\pi})^n}$). Аналогично, элементарному объекту $y_0$ разбиения $Q$ в пространстве $X$ соответствует функция $p(x) = \frac{1}{(\sqrt{2\pi})^n} e^{-iy_0 x}$ (при $y_0 = 0$  $p(x) = \frac{1}{(\sqrt{2\pi})^n}$). Следовательно, пространство $X$ не является базовым для объектов $q$, которые будут странными для него, а объекты $p$ будут странными для $Y$. Таким образом, пространства $X$ и $Y$ *несводимы*; объекты $p$ и $q$, несмотря на взаимно однозначное соответствие, являются *сторонними* друг другу, а представления $P$ и $Q$ имеют разную онтологию.

Преобразования Фурье лежат в основе *квантово-механического подхода* в физике, и релятивистские и квантово-механический подходы соответствуют несводимым представлениям, что во многом объясняет неудачи попыток объединения их в единую теорию, традиционно оперирующую в рамках одной системы представлений. Таким образом, квантово-механические объекты являются *странными* для релятивистских теорий, а их парадоксальные корпускулярно-волновые свойства связаны с различной онтологией. Дальнейший анализ даёт дополнительные подтверждения этому предположению.

## 3. Странные объекты в физике

В силу связности Сущего, все его части взаимодействуют между собой. Взаимодействие несводимых пространств будем называть их **сцеплением**, в общем случае, в некоторых **областях сцепления**. Результат сцепления будем называть **объектом сцепления**, который проявляет себя в сцеплённых пространствах как объект. Например, объекты $P$ и $Q$ на Рисунке 1Б являются результатом сцепления несводимых пространств $X$ и $Y$ в круговой области сцепления, которая также ограничивает область существования объектов $p$ и $q$.

Сцепление несводимых пространств $X$ и $Y$ в областях $X_Y \subseteq X$ и $Y_X \subseteq Y$ будем записывать как: $X \supseteq X_Y \rightleftharpoons Y_X \subseteq Y$. В общем случае возможно образование объектов сцепления двух $\{X_Y \rightleftharpoons Y_X\}$, трёх $\{X_{YZ} \rightleftharpoons Y_{XZ} \rightleftharpoons Z_{XY}\}$ и более пространств (см. Рисунок 2А). Если областью сцепления является всё пространство, то его будем называть **вложенным** в сцепленное с ним **вмещающее** пространство. На Рисунке 2Б пространство $Y$ вложено во вмещающее его пространство $X$: $X \supset X_Y \rightleftharpoons Y_X \equiv Y$; а $X$ вложено во вмещающее его пространство $Z$: $X \equiv X_Z \rightleftharpoons D_X^Z \subset Z$.

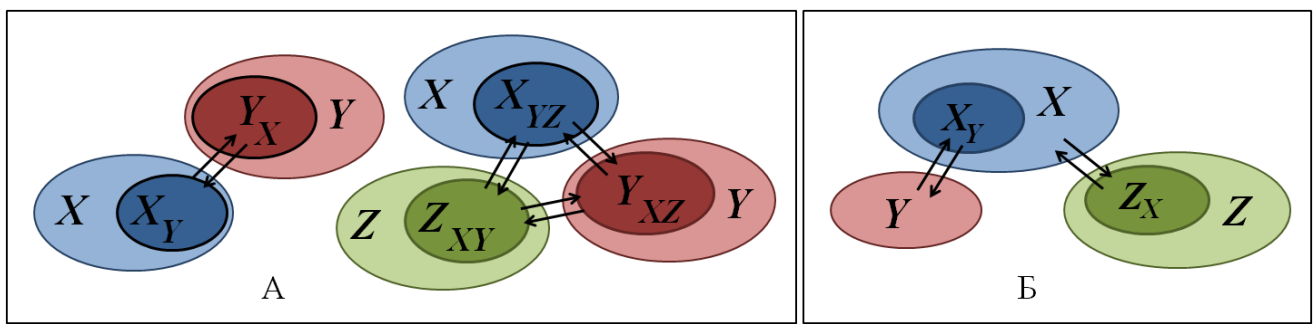

Рисунок 2. **Сцепления пространств (А), вложенные и вмещающие пространства (Б).**

### 3.1 Объекты квантовых теорий

В классических теориях (физика Ньютона, специальная и общая теории относительности) поведение физического объекта или системы описывается с помощью принципа наименьшего действия для интеграла: $S = \frac{1}{c}\int \Lambda \sqrt{-g}\,dX$, где $S$ – действие, $c$ – скорость света, $\Lambda$ – плотность функции Лагранжа системы, $dX$ – элемент объёма в пространстве-времени $X$ с метрическим тензором $g_{ik}$ и определителем g. В галилеевых координатах $g = -1$ и $S = \int L\,dt$, где $t$ – время, а $L = \int \Lambda\,dV$ – функция Лагранжа физической системы [8].

Если физический объект является результатом сцепления пространств в области $D \subset X$, то, выделяя эту область в действии $S$, получим:

$$S = S_f + S_m = \frac{1}{c}\int \Lambda_f \sqrt{-g}\,dX + \frac{1}{c}\int_D \Lambda_m \sqrt{-g}\,dX, \qquad (5)$$

где $S_m$ и $\Lambda_m$ описывают сам физический объект, а $S_f$ и $\Lambda_f$ – присущие ему взаимосвязи (поля) в пространстве $X$. В классических теориях физический объект рассматривается в рамках пространственно-временного континуума, то есть описывается как совокупность материальных точек. При этом функция $\Lambda_m$ приобретает смысл плотности распределения материальных точек в области $D$. Подобное представление соответствует замене странного объекта нормальным, что фактически выводит из рассмотрения процессы в областях сцепления. Как, например, замена странного объекта $Q$ из Примера 3 нормальным объектом $P$ с принципиально отличающейся внутренней структурой в области сцепления. Такое описание может быть удовлетворительным только вдали от области сцепления.

Вблизи и внутри области сцепления необходимо учитывать, что физические объекты $p, q, r, \ldots$ в сцепленных пространствах $X, Y, Z, \ldots$ являются проявлением объекта сцепления, состояние которого зависит от функций $p(x), q(y), r(z), \ldots$, определённых в соответствующих пространствах. Если ввести некоторые операторы $\hat{f}_{ij}$, описывающие взаимодействия сторонних объектов, то состояние объекта сцепления можно будет описать с помощью итерационных уравнений для вектора функций с параметром итерации $\xi$:

$$\begin{pmatrix} p_{\xi+1} \\ q_{\xi+1} \end{pmatrix} = \begin{pmatrix} \hat{f}_{11} & \hat{f}_{12} \\ \hat{f}_{21} & \hat{f}_{22} \end{pmatrix} \begin{pmatrix} p_\xi \\ q_\xi \end{pmatrix}, \quad \begin{pmatrix} p_{\xi+1} \\ q_{\xi+1} \\ r_{\xi+1} \end{pmatrix} = \begin{pmatrix} \hat{f}_{11} & \hat{f}_{12} & \hat{f}_{13} \\ \hat{f}_{21} & \hat{f}_{22} & \hat{f}_{23} \\ \hat{f}_{31} & \hat{f}_{32} & \hat{f}_{33} \end{pmatrix} \begin{pmatrix} p_\xi \\ q_\xi \\ r_\xi \end{pmatrix}, \ldots. \qquad (6)$$

Уравнения (6) переходят в обычные интегро-дифференциальные уравнения: $p_{\xi+1} = \hat{f} p_\xi$, аналогичные (5), если влияние сторонних объектов не учитывается. В этом случае параметр итерации $\xi$ будет представлять собой временные, пространственные координаты, интервал и т.п. в пространстве $X$.

В общем случае операторы взаимодействия $\hat{f}_{ij}$ неизвестны, но свойства физических объектов $p, q, r, \ldots$ накладывают ограничения на матрицу операторов $\hat{F}$. Например, условие стабильности объекта сцепления может быть задано уравнением:

$$\hat{F}^n \equiv \hat{I}, \qquad (7)$$

где $\hat{I}$ – тождественный оператор, соответствующий единичной матрице, а $n$ – количество итераций, необходимых для возвращения системы в исходное состояние: $(p_{\xi+n}, q_{\xi+n}, r_{\xi+n}, \ldots) \equiv (p_\xi, q_\xi, r_\xi, \ldots)$. Так, при сцеплении двух пространств необходимо два отображения: из $X$ в $Y$ и обратно для возвращения системы в исходное состояние (см. Рисунок 2А), и, согласно (7), условием стабильности физической системы будет: $\hat{F}^2 \equiv \hat{I}$. При сцеплении трёх пространств будет необходимо три отображения, и условием стабильности будет: $\hat{F}^3 \equiv \hat{I}$. При этом, возможно два варианта отображений: $X \to Y \to Z \to X$ и $X \to Z \to Y \to X$, что наблюдатель может интерпретировать как *спин* объекта.

В *квантовой механике* физическому объекту сопоставляется *волновая функция* $\Psi = \Psi_X + i\Psi_Y$, определённая в гильбертовом пространстве, что фактически соответствует описанию объекта сцепления двух пространств, характеризующемуся вектором состояния из двух функций (см. также Пример 4 предыдущего раздела). При этом условие стабильности объекта сцепления $\hat{F}^2 \equiv \hat{I}$ переходит в условие унитарности оператора $\hat{U}$, действующего на волновые функции: $\hat{U}^* \cdot \hat{U} \equiv \hat{I}$, где $\hat{U}^*$ – комплексно сопряжённый оператор. Для эрмитового оператора $\widehat{H}$, действующего на волновые функции в гильбертовом пространстве, унитарным будет оператор: $\widehat{U} = \exp(i\widehat{H})$. Таким образом, квантово-механический подход вполне может быть рассмотрен как частный случай сцепления пространств (6) с условием стабильности (7). Спин частиц описывается с помощью известных матриц Дирака.

Объекты сцепления двух и трёх пространств не могут быть представлены объектами сцепления меньшего количества пространств, поэтому не могут распадаться и будут стабильны, в отличие от объектов сцепления четырёх и более пространств, которые могут распадаться на объекты сцепления меньшего количества пространств, то есть могут быть **нестабильны**. Например, объект сцепления 4-х пространств может распадаться на два объекта сцепления 2-х пространств, а объект сцепления 5-ти пространств – на объекты сцепления 2-х и 3-х пространств. При этом, наряду с обычными законами сохранения, *спин*, как установившаяся последовательность взаимодействия сторонних пространств, также должен сохраняться. Примером *распада* объектов сцепления многих пространств в физике могут быть *резонансы*, *нестабильные частицы*, *ядра*. С объектами множественного сцепления приходится сталкиваться при *взаимодействии частиц высоких энергий*.

Такие наблюдаемые явления в физике элементарных частиц как парадоксальное удержание протонов и нейтронов в ядрах атомов (в КХД – *захват* кварков в адронах) вполне соответствуют модели множественных сцеплений, причём никаких дополнительных предположений о существовании новых фундаментальных *сильных* взаимодействий для 'удержания' сторонних объектов в области сцепления не требуется. Парадоксы возникают именно при попытке заменить сторонние объекты нормальными. Некоторые методы математического описания и квантования объектов сцепления предложены в [9].

Модель сцепления пространств не только даёт возможность общего описания физических объектов классических и квантовых теорий, но выявляет **причину** *появления* самих физических объектов как результат сцепления, что обычно постулируется как часть 'объективно существующего материального мира'.

### 3.2 Космологические объекты

В модель сцепления сторонних пространств вполне укладываются такие наблюдаемые космологические явления как аномальное притяжение видимого вещества галактик и ускоренное расширение вселенной, при этом никаких дополнительных предположений о 'неустранимой кривизне пространства' или о существовании **тёмной материи** и **тёмной энергии** неизвестной природы не требуется.

***Пример 5*** (*вложенные и вмещающие пространства*). Пусть координаты пространств $X$, $Y$ и $Z$ связаны соотношениями: $x = f(y)$ и $z = f(x)$, где $f$ – возрастающая, непрерывно дифференцируемая и ограниченная функция (например, $f(\xi) = \mathrm{arctg}(\xi)$, $f(\xi) = \xi/|\xi| \cdot [1 - \exp(-|\xi|)]$ и т.п.), то есть пространство $X$ вложено в $Z$ и вмещает $Y$: $Y \rightleftarrows X_Y \subset X \rightleftarrows Z_X \subset Z$ (см. Рисунок 2Б).

В силу условий, наложенных на функцию $f(\xi)$, её производная (в многомерном случае, градиент $\nabla f$) стремится к нулю $f'(\xi) \to 0$ при $\xi \to \infty$, а производная обратной функции на границе области сцепления, напротив, стремится к бесконечности $(f^{-1})'(\eta) \to \infty$. Таким образом, $\mathrm{d}x/\mathrm{d}y = f'(y) = 1/(f^{-1})'(x) \to 0$ на границе области $X_Y$, и $\mathrm{d}x/\mathrm{d}z = 1/f'(x) \to \infty$ при $x \to \infty$.

Поскольку между пространствами в областях сцепления установлено взаимно однозначное соответствие, сторонние объекты в этих областях могут быть идентифицированы, при этом их поведение будет отличаться от поведения нормальных объектов, что будет выглядеть странным и необъяснимым для наблюдателя. Так, вместо ожидаемой скорости $v_y = \mathrm{d}y(t)/\mathrm{d}t$, где $t$ – время в пространстве $X$, объектов вложенного пространства $Y$ в пространстве $X$, их наблюдаемая скорость $v_x = \mathrm{d}x(t)/\mathrm{d}t$ при приближении к границам области сцепления $X_Y$ будет стремиться к нулю: $v_x = v_y/(f^{-1})'(x) \to 0$, как если бы на них действовали невидимые силы, удерживающие их внутри области сцепления. Подобный эффект, необъяснимый для нормальных объектов, наблюдается в космологии в виде аномального удержания видимого вещества галактик, что 'объясняется' гравитационным притяжением к невидимой ***тёмной материи*** неизвестной природы.

Видимая скорость $v_x$ объектов вмещающего пространства $Z$ в пространстве $X$, напротив, будет увеличиваться по сравнению с ожидаемой $v_z = \mathrm{d}z(t)/\mathrm{d}t$ при $x \to \infty$: $v_x = v_z/f'(x) \to \infty$, то есть в пространстве $X$ объекты вмещающего пространства будут наблюдаться как разлетающиеся с ускорением. Подобный эффект в космологии интерпретируется как ускоренное расширение вселенной.

Формально, расширение вселенной можно описать и в рамках общей теории относительности путём добавления известного космологического члена $\Lambda_0 g_{ik}$ в уравнения Эйнштейна для гравитационного поля. Однако, появление этого члена в рамках общей теории относительности необъяснимо, так как означает *"приписывание пространству-времени неустранимой кривизны, не связанной ни с материей, ни с гравитационными волнами"* [8]. Сопоставление космологического члена с тензором энергии-импульса $T_{ik}^{DE}$ некой невидимой ***тёмной энергии*** неизвестной природы, равномерно распределённой во всём пространстве, приводит к равенству: $\frac{8\pi k}{c^4} T_{ik}^{DE} = \Lambda_0 g_{ik}$, что при $g_{ik} = \mathrm{diag}\{1,-1,-1,-1\}$ соответствует: $T_{ik}^{DE} = \mathrm{diag}\{\rho,-\rho,-\rho,-\rho\}$, где $\rho = \Lambda_0 c^4/8\pi k$, то есть при положительной плотности энергии введённой тёмной энергии её давление и импульс будут отрицательными. В отличие от 'необъяснимой кривизны' пространства или неизвестной тёмной энергии с нефизическими свойствами, поведение сторонних объектов вмещающего пространства имеет ясный физический смысл и не требует дополнительных интерпретаций. Неопределённости, парадоксы и противоречия возникают именно при попытке приписать странным объектам свойства нормальных.

### 3.3 Формирование вселенной и физический вакуум

Принципы свободы при формировании вселенной В. Докучаев в [10] анализирует на основе сопоставления с возможными 'способами существования белковых тел':

- 'английский': разрешено всё, что не запрещено;
- 'немецкий': разрешено только то, что не запрещено;
- 'французский': разрешено всё и даже то, что запрещено;
- 'китайский': запрещено всё и даже то, что разрешено;

и делает вывод о том, что принцип свободы при формировании вселенной соответствует 'английскому': *разрешено всё, что не запрещено*. В отличие от вселенной, Сущее включает в себя различные по онтологии системы представлений и присущие им взаимные противоречия и парадоксы, поэтому принцип свободы при его формировании, назовём его 'русский', должен соответствовать компиляции всех рассмотренных выше принципов, где разрешённое относится в основном к сторонним системам, а запрещённое – к собственным:

- 'русский': разрешено всё, что не запрещено, и даже то, что запрещено, но не в собственной системе, где разрешено только то, что не запрещено, и запрещено даже то, что разрешено.

В этой связи принципиальное значение приобретает вопрос: Все ли из возможных разбиений могут существовать, и если нет, то какие именно? Наиболее последовательным с точки зрения сформулированного принципа свободы выглядит предположение о существовании всевозможных разбиений, но субъект будет воспринимать только те, результат сцепления с которыми в собственном пространстве он сможет идентифицировать. Остальные разбиения наблюдатель будет 'воспринимать' как влияние чего-то неопределённого, того, что не может быть опознано им как объекты. В основном, эти влияния будут хаотичны и разнонаправлены, поэтому будут компенсировать друг друга, как, например, проявления виртуальных частиц *физического вакуума*. Если воздействия, случайно или нет, окажутся согласованы, то наблюдатель вполне сможет их идентифицировать. Например, в виде *возникновения частиц* из физического вакуума вблизи гравитационного радиуса чёрных дыр или как *рождение вселенной* в результате случайных флуктуаций физического вакуума.

Внешние зоны областей сцепления вмещающих пространств (область $Z \setminus Z_X$ на Рисунке 2Б) кажутся недоступными для наблюдателей из вложенных пространств, что на первый взгляд противоречит связности Сущего. Это кажущееся противоречие имеет вполне ясное толкование: все части Сущего взаимосвязаны, но взаимосвязи с внешними областями вмещающих пространств недоступны для систем представлений вложенных пространств, так как выходят за рамки, определяемые их онтологией.

Недоступность представлений, выходящих за рамки собственной системы, вплотную приводит к проблеме понимания в познании.

### 4. Проблема понимания в науке

Традиционно проблема понимания тесно связана с наличием *непротиворечивой* системы представлений, в науке – стройной теории, которая может описать все наблюдаемые явления окружающего мира. То, что не укладывается или выходит за рамки системы, воспринимается как находящееся за гранью ***понимания***, поэтому стремление А. Эйнштейна 'свести в систему' соответствует желанию понять 'все наблюдаемые явления' в окружающем мире. Однако, основателю релятивистской картины мира так и не удалось сделать это. Камнем преткновения оказалась зарождающаяся в то время квантовая механика с её странными дуальными объектами, неопределённостями и парадоксами. *"Бог не играет в кости"*, – говорил А. Эйнштейн, образно указывая на то, что понятия и аппарат квантовой механики не просто несовместимы с подходом релятивистских теорий, а нарушают принципы построения релятивистской картины мира, на что, правда, получил не менее образное возражение Н. Бора: *"Не указывайте Богу, что ему делать"*. Как бы там ни было, но, несмотря на множество попыток, 'свести в систему' релятивистские и квантовые теории так и не удалось. Квантовая механика остаётся "*полной загадок и парадоксов дисциплиной, которую мы не понимаем до конца, но умеем применять*" (М. Гелл-Манн [11]), дисциплиной, которую "*не понимает никто*" (Р. Фейнман [5]).

Не менее странными объектами являются и 'открытые' относительно недавно тёмная материя и тёмная энергия. Природа их неизвестна, и введены они были для объяснения наблюдаемого аномального удержания видимого вещества галактик и ускоренного расширения вселенной соответственно. Впечатляет тот факт, что, согласно оценкам, тёмная материя и энергия неизвестной природы составляют более 95 % вещества во вселенной, что характеризует масштаб непознанного, вероятно, далеко не окончательный.

Но и на этом проблемы понимания в науке не заканчиваются: системный подход встречает принципиальные возражения, причём со стороны самого системного подхода. В математической теории множеств это выражается в системных парадоксах Кантора и Рассела [12-15]. *Парадокс Кантора* ставит под сомнение возможность построения наиболее общего множества всех множеств (то есть единой непротиворечивой и всеобъемлющей системы представлений), поскольку мощность этого множества всегда будет меньше мощности множества всех его подмножеств. *Антиномия Рассела* подвергает сомнению возможность классификации и систематизации с помощью множеств (и систем представлений) вообще. Действительно, если определить множество, не содержащее себя в качестве элемента, как 'обычное', а 'необычным' считать множество, содержащее себя в качестве элемента, то множество Рассела – множество 'обычных' множеств не может быть отнесено ни к обычным, ни к необычным. Ведь если оно обычное, то содержит себя в качестве элемента и, следовательно, является необычным, а если оно необычное, то не содержит себя в качестве элемента и является обычным. В обоих случаях мы приходим к противоречию, что говорит о том, что множество Рассела не поддаётся классификации и систематизации.

Общепризнанного пути преодоления системных парадоксов теории множеств не существует. Предлагаемые способы устранения парадоксов путём накладывания тех или иных ограничений на множества или операции с ними, с точки зрения физики выглядят как очередные попытки 'указать Богу, что ему делать'. Кроме того, неопределённости и парадоксы проявляются в свойствах наблюдаемых физических объектов, поэтому из наиболее общей системы устранены быть не могут. Наука вплотную подошла к принципиальному противоречию: системный подход требует устранения неопределённостей и парадоксов, но они являются неотъемлемой частью окружающего мира, то есть требуется устранить то, что устранить невозможно. Именно это противоречие отмечалось во введении как кризис системного подхода в научном познании.

Многократные безуспешные попытки объединения релятивистских и квантовых теорий, 'открытие' космологических объектов с нефизическими свойствами, непреодолимые системные парадоксы привели к установлению новой научной парадигмы: к *математическому формализму*. Апологеты этой парадигмы утверждают, что наука вторглась в области настолько далёкие от житейских, что сопоставить открываемые объекты с существующими представлениями невозможно, поэтому остаётся описывать явления математически с помощью формул, не вдаваясь в смысл вводимых понятий, а то и просто не вводя никаких понятий. При этом апологеты математического формализма на высказывания, что '*квантовую механику не понимает никто*', отвечают словами Дирака: "*Тут нечего понимать!*", и приводят слова Галилео Галилея утверждавшего, что '*Природа написана на языке математики, и мы должны изучать этот язык*'.

Абсурдность аргументов заключается в том, что слова Галилея были обращены к его оппонентам, считавшим, помимо всего прочего, что гелиоцентрическая модель, даже подкреплённая наблюдениями и расчётами Галилея, вовсе не нужна, так как определить положение планет вполне можно по таблицам, а что там вокруг чего вращается – это совершенно неважно, тем более, что вряд ли когда-нибудь появится возможность проверить это. Таким образом, Галилей выступал именно за *понимание* законов природы и законов движения планет в частности, а его оппоненты – за формальное применение таблиц. Нынешние формалисты также выступают за формальное использование математических формул, при этом ещё и прикрываясь именем Галилея...

Возникает парадоксальная и опасная ситуация, когда наука в лице апологетов математического формализма отрекается от своих главных достоинств – систематизации и понимания, и эта тенденция становится доминирующей *парадигмой*, препятствующей развитию науки. Согласно теории Томаса Куна [16], из-за активного противодействия апологетов доминирующей парадигмы изменить её очень непросто, и чаще всего изменение или смена парадигмы происходит посредством научных революций. Теория Т. Куна подверглась острой критике со стороны ряда учёных 'за непонимание принципиального значения принципа соответствия между старыми и новыми теориями, за отсутствие подлинного историзма, за непонимание неоднородности развития науки' и т.п. (В.Л. Гинзбург [17]). Однако, впоследствии, отдельные критики, в полном соответствии с положениями теории Куна, создали так называемую 'Комиссию по *борьбе* (!) с лженаукой'. Нетрудно догадаться, что для апологетов доминирующей парадигмы лженаукой будет являться всё, что выходит за рамки этой парадигмы. А ведь в своё время гелиоцентрические взгляды Галилея также были провозглашены еретическими теми, кто считал себя единственными блюстителями Правды и Знания, и наука на несколько веков была отброшена назад...

Казалось бы, о каком системном подходе и понимании может идти речь в *концепции несводимых представлений*, если несводимые представления принципиально не могут быть объединены в рамках одной системы, а ведь именно невозможность создания единой теории привело к отсутствию понимания и послужило началом математического формализма? Кроме того, концепция не предлагает способ решения принципиальных системных парадоксов теории множеств, а лишь подтверждает невозможность построения наиболее общей системы.

Всё это так, и, тем не менее, концепция несводимых представлений позволяет решить все эти проблемы, хоть и несколько неожиданным способом. Во-первых, разграничивая несводимые системы, концепция позволяет вывести за рамки систем неустранимые противоречия и парадоксы, определяемые разницей их онтологий, что сохраняет внутреннюю непротиворечивость систем и *понимание* в рамках каждой из них. При этом, несводимость систем является необходимым условием, так как в противном случае противоречия могли бы быть устранены. Таким образом, концепция даёт принципиальную возможность решить проблему понимания в науке и преодолеть кризис системного подхода в научном познании.

Во-вторых, существование множества несводимых систем означает существование различных законов и логик в них, поэтому можно вполне обоснованно использовать *различные* методы преодоления системных парадоксов, в том числе и из теории множеств (более детально см. [18]), при этом 'не указывая Богу, что ему делать', так как существование любой системы и логики не отрицает существования других, даже противоречащих ей.

В этой связи особую важность приобретают теоретические исследования взаимодействия несводимых систем и противоречащих друг другу логик. Являясь частями единого и связного Сущего, несводимые системы представлений воздействуют и формируют друг друга, что определяет **эволюцию** Сущего. Принцип свободы не ограничивает вид несводимых систем, но этот вид будет определять результат их взаимодействия. Например, согласованное сцепление пространств может привести к образованию физических частиц и вселенной, а несогласованное – к образованию виртуальных частиц физического вакуума.

Фактически, концепция несводимых представлений является Единой теорией, объединяющей релятивистские и квантовые теории, странные объекты космологии, только объединение это происходит на основе обобщённого метода познания за онтологическими рамками систем представлений, когда пропадают сами понятия объектов и систем. С точки зрения традиционного метода познания, объединение происходит путём разделения онтологически несовместимых систем.

Таким образом, концепция несводимых представлений значительно расширяет возможности научного познания, распространяя его с одной на множество различных по онтологии систем представлений, при этом снимая принципиальные противоречия системного подхода в научном познании и объединяя множество несовместимых подходов в многополярную научную картину мира.

## 5. Многополярная картина мира

Обобщённый подход к познанию по определению лежит в основе любых представлений человека: от бытовых и религиозных представлений до научных теорий, философских систем и мировоззрения в целом. Сущее содержит различные системы, включая несводимые, и, как следствие, содержит неопределённости, парадоксы и противоречия между ними.

Например, *материальный мир*, понимаемый в материалистическом подходе как некая 'объективная реальность, данная в ощущениях', влияет на человека как минимум через ощущения и по определению является частью Сущего. *Субъективные представления*, понимаемые как образы материальных объектов, отличаются 'по природе' от познаваемых объектов материального мира, но тоже оказывают влияние на человека, определяя его поведение, и, следовательно, также являются частью Сущего.

Представления, понятия и идеи, принятые в обществе – *общие понятия*, такие как язык общения, законы, моральные и нравственные нормы, научные и религиозные представления и т.п., также оказывают влияние на человека и являются частью Сущего. При этом они отличаются как от материальных объектов, поскольку являются представлениями, так и от субъективных представлений, поскольку непосредственно к субъекту не относятся и являются объективными для него. По сути, общие понятия являются продуктом или сублимацией субъективных представлений *многих* познающих, что и определяет их объективный характер для каждого субъекта в отдельности.

В материалистическом подходе относительная объективность общих понятий отличается от абсолютной объективности материального мира, вообще независимой от познающего субъекта. Однако, познающие не могут идентифицировать сторонние объекты и познающих с различной онтологией, поэтому их воздействия будут восприниматься как нечто объективное. Можно предположить, что, как и сублимация субъективных представлений в общепринятые, сублимация субъективных и общепринятых представлений систем с различной онтологией будет составлять некий абсолютно объективный *идеальный мир*, что является исходным пунктом *объективного идеализма*.

Таким образом, *концепция несводимых представлений* выявляет онтологическую суть разделения таких философских направлений как материализм, субъективный и объективный идеализм и связывает их воедино как взаимодействующие, взаимно дополняющие и формирующие друг друга несводимые системы представлений. Так, окружающий нас *материальный мир* формирует *субъективные представления*, из которых сублимируются *общие*, так же как из различных по онтологии *понятий* складывается *идеальный мир*, являющийся источником *объективных физических закономерностей* и *объектов* в каждом из различных по онтологии окружающих миров, в частности, нашего материального мира. Подобные ***циклические цепочки несводимых сущностей***, взаимодействующие и формирующие друг друга, определяют ***эволюцию*** Сущего, а ***противоречия*** между несводимыми сущностями выступают в качестве движущей силы эволюции.

С этой точки зрения становится очевидным ограниченность традиционного в *естественных науках* материалистического подхода, в основном оперирующего в рамках одной онтологической системы представлений, из которой диалектическое единство несводимых систем воспринимается как парадоксальное, полное неопределённостей и противоречий. Концепция несводимых представлений позволяет выйти за пределы онтологической системы и решить многие принципиальные проблемы системного подхода в научном познании, попутно соединяя многие научные теории и подходы. Так, например, сделанный ранее вывод о том, что 'природа' познающего субъекта определяет онтологию воспринимаемого им окружающего мира, вполне может быть обоснованием антропного принципа [19,20] формирования вселенной. А фрактальные отображения Б. Мандельброта [21,22], являющиеся основой теории диссипативных структур И. Пригожина [23,24], могут быть рассмотрены как частный случай итерационных отображений сцеплённых пространств.

Ограничиваясь изучением исключительно материального мира и устраняя *противоречия* из соответствующей системы представлений, материалистический подход выводит из рассмотрения основные движущие силы эволюции Сущего и материального мира в частности. В этой связи наивно выглядят попытки объяснить происхождение материального мира, оставаясь в рамках материалистического подхода, фактически, выводящего из рассмотрения организующее начало в Сущем. Концепция несводимых разбиений позволяет выйти за рамки материалистического подхода и выявить движущие силы эволюции, необходимые для обоснования принципа самоорганизации материи в синергетических теориях, таких, например, как теория самоорганизации вселенной В. Бранского [25].

Научный подход обоснованно критикует *религиозные системы* за противоречивость и бездоказательность, но именно эти недостатки позволили им выйти за рамки материалистического подхода, и приблизиться к восприятию Сущего в динамике, вместе с присущими ему противоречиями и парадоксами. Так в *индуизме* Триединое божество – Тримурти (в переводе с санскрита: *три лика*) объединяет три главные божества индуистского пантеона – Брахму-Создателя, Вишну-Хранителя и Шиву-Разрушителя, которые соответствуют созидающему, организующему и разрушающему началам в Сущем, определяющим его самоорганизацию и эволюцию. Подобные силы созидания, сохранения и разрушения лежат в основе *буддийского круга Сансары*, соответствующего рассмотренному ранее *циклу эволюции*, повторяющемуся в разных вариациях на разных уровнях или частях Сущего. В *христианстве* единство Сущего отражается не в виде единства противодействующих сущностей-сил, а в виде единства несводимых сущностей-форм различной природы. Так, центральное понятие Святой Троицы персонифицируется в виде Триединства Бога-Отца, Бога-Сына и Святого Духа, имеющих явные соответствия с понятиями идеального (абсолютный мир идей), материального (материальное воплощение Бога-Отца) и субъективного (движущая сила и основа формирования) соответственно. Взаимодействие этих сущностей было рассмотрено ранее при анализе философских направлений. Аналогии с несводимыми представлениями позволяют трактовать единство Бога не в плане *монотеизма*, подразумевающего единую систему, единоначалие и *единообразие* мира, а как единство мира во всём *многообразии* взаимодействующих и формирующих друг друга несводимых сущностей, что принципиально противоположно единообразию.

В *социальном* плане единство многообразия соответствует концепции *многополярного мира*, понимаемого как единство *различных* социальных систем, взаимосвязанных, взаимодействующих, дополняющих и формирующих друг друга. Распространённой, но принципиальной ошибкой является понимание многополярного мира как множества одинаковых 'полюсов', то есть одинаковых социальных систем, что, как было показано в работе [26], приводит к конкурентной борьбе и представляет собой один из этапов глобализации на пути к единоначалию и единообразию, противоположному многообразию. Многообразие подразумевает наличие несводимых, то есть принципиально несовместимых социальных систем, что исключает их слияние, таких, например, как монархизм, капитализм и социализм. Взаимосвязь несводимых социальных систем определяет их взаимное развитие, а попытки их механического объединения путём искусственного смешения людей и внедрения 'толерантности', напротив, приведут лишь к обострению непреодолимых противоречий, к конфликтам и общей деградации насильственно 'объединяемых' таким образом социальных систем, что и наблюдается в различных регионах современного мира, где пытаются утвердить глобализацию.

Единое, замкнутое и связное Сущее объединяет множество несводимых систем представлений с различной онтологией, соответствующих множеству окружающих миров. Онтологическая несовместимость и несводимость систем и миров не означает их изолированность друг от друга. Напротив, являясь частями единого Сущего, они тесно взаимосвязаны, взаимно дополняют, обогащают, формируют и совершенствуют друг друга, при этом не смешиваясь и оставаясь различными 'по природе'. Возникающие циклические цепочки взаимосвязанных несводимых сущностей, образующиеся на разных уровнях Сущего, определяют его эволюцию, а противоречия между сущностями выступают в качестве движущей силы эволюции.

## Заключение

Введённый постулат существования и познаваемости позволяет уточнить многие понятия теории познания, разграничить непознанное и непознаваемое, обобщить процесс познания Сущего на всевозможные взаимодействия его частей, что выявляет ограниченность традиционного процесса познания, заключающемся в идентификации объектов окружающего мира, природой или онтологией познающего субъекта. Таким образом, обобщённый процесс познания приводит к концепции несводимых представлений – к возможности существования различных по онтологии систем представлений, несовместимых друг с другом на уровне понятий и определений, поэтому несводимых в рамках одной непротиворечивой системы.

Разделяя несовместимые системы представлений, концепция несводимых представлений устраняет внутренние противоречия в каждой из систем, чем решает проблему системного подхода в научном познании и связанную с ним проблему понимания в науке. Фактически, концепция несводимых представлений является Единой теорией, объединяющей релятивистские и квантовые теории, странные объекты космологии, только объединение это происходит на основе обобщённого метода познания, то есть за онтологическими рамками систем представлений, когда пропадают сами понятия объектов и систем. С точки зрения традиционного метода познания, объединение происходит путём разделения онтологически несовместимых систем.

Единое, замкнутое и связное Сущее объединяет множество несводимых систем представлений с различной онтологией, соответствующих множеству окружающих миров. Являясь частями единого Сущего, несовместимые и несводимые системы и миры тесно взаимосвязаны, взаимно дополняют, обогащают, формируют и совершенствуют друг друга, при этом, не смешиваясь и оставаясь различными по природе. Возникающие циклические цепочки взаимосвязанных несводимых сущностей, образующиеся на разных уровнях Сущего, определяют его эволюцию, а противоречия между сущностями выступают в качестве движущих сил эволюции.

Концепция несводимых представлений значительно расширяет возможности научного познания, распространяя его с одной на множество различных по онтологии систем представлений, при этом снимая принципиальные противоречия системного подхода в научном познании и объединяя множество несовместимых подходов. Обобщённый подход к познанию по определению лежит в основе различных систем представлений: от бытовых и религиозных до научных теорий, философских систем и мировоззрения в целом, что даёт возможность общего анализа и уточнения их основ, приводит к многополярной картине мира.

## Список литературы